\documentclass[sigconf]{acmart}
\renewcommand\footnotetextcopyrightpermission[1]{} 
\AtBeginDocument{%
  }

\setcopyright{acmlicensed}
\copyrightyear{2026}
\acmYear{2026}
\acmDOI{XXXXXXX.XXXXXXX}
\acmConference[Preprint]{Preprint}{}{}
\acmISBN{978-1-4503-XXXX-X/2026/06}

\usepackage{subcaption}
\usepackage{booktabs}
\usepackage{multirow}
\usepackage{adjustbox}
\usepackage{enumitem}
\usepackage{array}

\begin{document}

\title{Mirage: Transmitting a Video as a Perceptual Illusion for 50,000x Speedup}



\author{Junjie Wu}
\affiliation{%
  \institution{Southwest Jiaotong University}
  \city{Chengdu}
  \state{Sichuan}
  \country{China}
}
\email{jjw@swjtu.edu.cn}

\author{Tianrui Li}
\affiliation{%
  \institution{Southwest Jiaotong University}
  \city{Chengdu}
  \country{China}}
 \email{trli@swjtu.edu.cn}

\author{Yi Zhang}
\affiliation{%
  \institution{Sichuan University}
  \city{Chengdu}
  \country{China}
}   \email{yzhang@scu.edu.cn}

\author{Ziyuan Yang}
\authornote{
Ziyuan Yang is the corresponding author.}
\affiliation{%
  \institution{Sichuan University}
  \city{Chengdu}
  \country{China}}
\email{cziyuanyang@gmail.com}







\begin{abstract}

The existing communication framework mainly aims at accurate reconstruction of source signals to ensure reliable transmission. However, this signal-level fidelity–oriented design often incurs high communication overhead and system complexity, particularly in video communication scenarios where mainstream frameworks rely on transmitting visual data itself, resulting in significant bandwidth consumption. To address this issue, we propose a visual data-free communication framework, \textit{Mirage}, for extremely efficient video transmission while preserving semantic information. Mirage decomposes video content into two complementary components: temporal sequence information capturing motion dynamics and spatial appearance representations describing overall visual structure. Temporal information is preserved through video captioning, while key frames are encoded into compact semantic representations for spatial appearance. These representations are transmitted to the receiver, where videos are synthesized using generative video models. Since no raw visual data is transmitted, Mirage is inherently privacy-preserving.

Mirage also supports personalized adaptation across deployment scenarios. The sender, network, and receiver can independently impose constraints on semantic representation, transmission, and generation, enabling flexible trade-offs between efficiency, privacy, control, and perceptual quality. Experimental results in video transmission demonstrate that Mirage achieves up to a $50{,}000\times$ data-level compression speedup over raw video transmission, with gains expected to scale with larger video content sizes.

\end{abstract}

\ccsdesc[500]{Networks~Network architectures}
\ccsdesc[300]{Networks~Network protocols}
\ccsdesc[300]{Computing methodologies~Computer vision}
\ccsdesc[300]{Computing methodologies~Machine learning}

\keywords{Semantic Communication; Generative Video Transmission; Wireless Networks; Video Understanding; Personalized Communication}



\maketitle



\section{Introduction}
Existing communication framework are largely grounded in Shannon’s information-theoretic framework, where the primary objective is to enable reliable transmission and high-fidelity reconstruction of source signals at the receiver. In this framework, communication systems operate on signal representations and ensure bit-level reliability through source and channel coding mechanisms~\cite{Wei10740493,ieee10217138}.

However, this fidelity-oriented transmission paradigm suffers from a fundamental limitation:
its resource consumption is increasingly unsustainable under limited spectrum resources. Video data, as one of the dominant bandwidth consumers, exemplifies this challenge.
For large-scale video signals, traditional compression methods exploit spatial and temporal
statistical correlations at the pixel level. Nevertheless, compression efficiency is ultimately
constrained by channel conditions. 
Hence, in video transmission, limited bandwidth can only be accommodated by sacrificing
resolution or frame rate, which inevitably degrades semantic fidelity.
This limitation is even more pronounced in video communication, where both spatial appearance
and temporal dynamics must be preserved simultaneously, resulting in significantly higher
bandwidth requirements for reliable pixel-level transmission.

To address these challenges, semantic communication has emerged as a promising paradigm.
Semantic communication, as conceptualized in Weaver's framework~\cite{Weaver2009RecentCT}, is concerned with the accurate conveyance of meaningfrom the sender to the receiver, ensuring that the transmitted information is not only received without error but also correctly understood and actionable.
Recently, semantic communication has been actively explored in image and text transmission.
For example, Dai \emph{et al.}~\cite{Dai2022} proposed NTSCC, which learns latent semantic features
and entropy priors to optimize perceptual rate–distortion performance.
To adapt to dynamic data environments,  Zhang \emph{et al.}~\cite{semantic1} further introduced receiver-guided semantic encoding
with data-adaptive mechanisms.
Peng \emph{et al.}~\cite{peng12024} studied robustness in semantic communication and proposed the R-DeepSC
architecture to mitigate semantic distortion during transmission.
The key idea of semantic communication is to transmit only task-relevant semantic information,
thereby eliminating task-irrelevant redundancy at the source~\cite{weng2021semantic,Dai2022,zhang2023}.
By moving beyond bit-level transmission and directly encoding meaning, this paradigm enables
orders-of-magnitude improvements in communication efficiency.
For static sources such as images, semantic communication has demonstrated strong potential,
since semantic information is primarily embedded in two-dimensional spatial structures.

Extending semantic communication to video transmission, however, introduces a fundamental
challenge: transitioning from \emph{static semantics} to \emph{spatiotemporal semantics}.
Video semantics arise from the joint interaction of spatial appearance and temporal dynamics.
The high computational complexity and temporal uncertainty of video signals prevent direct
application of existing semantic communication frameworks.
This high-dimensional spatiotemporal coupling creates a dilemma:
extreme compression is required to reduce bandwidth consumption,
yet excessive compression damages video quality or semantic integrity, causing semantic inconsistency
or deviation from the original intent at the receiver~\cite{10631278}.
Hence, existing solutions fail to reconcile the tension between
\emph{high-dimensional semantic representation} and \emph{efficient transmission} for video.

Revisiting this situation, we find that suppressing errors in high-fidelity visual signals transmitted through noisy channels may require higher quantization bit depths. This leads to a further increase in the amount of data to be transmitted, inevitably exacerbating channel occupancy time and placing even greater pressure on already scarce spectrum resources. This motivates a new question:

\noindent \textbf{\emph{``Can video communication be achieved without transmitting raw visual data directly?"}}

Video can be decomposed into two complementary information: a) \textit{Spatial Information}: depicting static structures; b) \textit{Temporal Information}: representing dynamic processes. Hence, in this paper, we aim to compactly condense both information to reduce communication overhead.

In this paper, we reformulate the video transmission problem as a combination of the video understanding task and the representation task, and propose a novel visual-data-free video transmission framework \textit{Mirage}.
However, at the receiving end, simply decoding the semantic representation of the video is insufficient, as most downstream intelligent tasks and the receiver itself require visualized visual data operating in the pixel domain. A crucial "semantic-to-visual" gap remains between semantic representation and the final usable video data.
Motivated by recent advances in text-to-video generation~\cite{NEURIPS2023_52f05049,ICCV2023}, we replace the conventional signal decoding process with a generative reconstruction process, which is guided by video understanding and compact representation features.

Specifically, \textit{Mirage} consists of three main steps, including video understanding, semantic communication, and video generation. First, to reduce the overhead of temporal information, we designed a captioner and text encoder on the sender side. The captioner captures the temporal information representation of the video through video captioning, and the text encoder further compresses the video caption losslessly to minimize the amount of information used to describe the dynamics. Then, to reduce spatial information, we designed a selector and a semantic encoder. The selector selects the most task-oriented keyframe by aligning the sender's or receiver's intent, and the semantic encoder extracts compact representation features of the keyframe, achieving spatial information compression in both temporal and spatial dimensions.

However, complex channel conditions in the semantic communication process can cause information errors, posing a major threat to visual spatial information that requires high-fidelity transmission. Mirage's spatiotemporal decoupling design addresses this challenge by providing dual transmission guaranties: highly robust text instructions can resist channel interference, while the keyframe stream can be moderately degraded under harsh channel conditions to ensure spatial semantic consistency. This jointly guaranties reliable transmission of video semantics under dynamic channel conditions.

Finally, unlike previous decoding-based works, Mirage performs "generation" rather than "decoding." Specifically, spatial information provides visual anchors for the generator, while temporal information acts as conditional features to guide the generator's noise generation. The generator, through conditional denoising, can reconstruct a video with a certain semantic consistency to the original video. We refer to the video spatial and temporal information under the Mirage architecture as "\emph{perceptual illusion}." Moreover, the entire process only transmits the "\emph{perceptual illusion}" of the video decomposition without transmitting the original data, making it more privacy-preserving compared to traditional methods.

Notably, benefiting from transferring the encoding process into the generation process, our proposed \textit{Mirage} has a unique and interesting characteristic, \textit{personalization}. 
The entire communication process involves three participants: the sender, the network, and the receiver. Each party can personalize the transmitted content within predefined limits according to their own needs. Specifically, at the sending end, personalization influences text prompts by adjusting semantic granularity and content, and determines the visual semantic anchors used for communication transmission by adjusting keyframe selection. The network adaptively performs content moderation and semantic transmission strategy selection based on channel conditions and network utility objectives, which is reflected in the adjustment of communication parameters such as codebook size, image resolution, and compression ratio. As the recipient, the receiver further personalizes the perceived results through generation control, including visual enhancement, semantic prompt adaptation, and conditional generation, thereby achieving personalized video generation that meets expectations such as visual identity consistency ("looks like me"), semantic applicability ("content is suitable for me"), and preferred interaction style ("expressed in a way I like"). The Mirage, which is designed to explicitly model the personalization needs of all three participants, can achieve these three targets separately.


The main contributions of this work are summarized as follows:
\begin{itemize}[leftmargin=*]
    
    \item We propose Mirage, a visual data-free communication framework that replaces pixel-level video transmission with compact semantic representations and receiver-side content synthesis, enabling video communication without delivering raw or compressed visual signals.
    
    \item We design an end-to-end system architecture that structurally separates semantic representation, network transmission, and content generation, allowing sender-, network-, and receiver-specific constraints to be independently expressed and jointly satisfied.
    
    \item We implement a Mirage pipeline and conduct extensive evaluations on video transmission tasks, demonstrating up to a $50{,}000\times$ data-level compression speedup over raw video transmission while maintaining strong semantic consistency and perceptual quality.
\end{itemize}

\section{Background}

\subsection{Semantic Communication}

Semantic communication \cite{Lan9663101} has been proposed as a promising alternative to conventional data-centric communication, shifting the focus from bit-level fidelity to task-oriented semantic correctness. Instead of transmitting raw signals, semantic communication systems aim to extract task-relevant features or representations and deliver them to the receiver, often using learned joint source--channel coding schemes. This paradigm has demonstrated significant gains in communication efficiency and robustness, especially in bandwidth-constrained or noisy channel conditions.

Existing research on semantic communication has focused primarily on improving transmission efficiency for specific downstream tasks, such as image classification \cite{weng2021semantic,Dai2022,zhang2023} or text delivery \cite{peng12024,Yan22022}. More recently, attention has expanded to video-centric scenarios, including video question answering based on VideoQA \cite{guo2025}, world models for video frame prediction \cite{jiang2025semantic}, and semantic-aware adaptive video transmission schemes \cite{Yan12025}. By prioritizing task-relevant semantic features and allocating communication resources accordingly, these approaches effectively reduce redundant information transmission and improve task performance. However, despite their efficiency gains, most existing semantic communication methods remain fundamentally reconstruction-oriented: the receiver is still required to recover the transmitted semantic signals or feature representations, rather than directly satisfying semantic or perceptual objectives.


\subsection{Video Understanding and Generation}
Video understanding aims to extract high-level semantic information from raw video streams, enabling machines to reason about visual content beyond pixel-level representations \cite{Abdar2024}. Among various video understanding tasks, video captioning has emerged as a representative approach that distills complex spatio-temporal visual signals into compact, human-interpretable textual descriptions \cite{NEURIPS2024_22a7476e,Liang2022,yang2025}. Recent advances have shown that high-quality caption generation requires not only dense temporal modeling but also fine-grained spatial understanding and scalability to long videos. For example, ShareGPT4Video constructs large-scale, densely annotated video-caption pairs to improve temporal coherence and caption precision without relying on expensive manual labeling, addressing challenges in temporal alignment, content completeness, and long-video scalability \cite{NEURIPS2024_22a7476e}. 

Beyond data-centric approaches, prior work has explored structured relational modeling for video captioning. Liang et al. propose a long short-term graph (LSTG) to jointly capture short-term spatial semantic relations and long-term temporal dependencies among objects, and introduce a global gated graph reasoning module to mitigate relation ambiguity and over-smoothing, leading to improved caption generation performance \cite{Liang2022}. In parallel, recent large-scale multi-modal models further enhance caption richness by tightly integrating visual, temporal, and linguistic representations. For instance, DAM \cite{dam} introduces localized visual encoding and focus prompting to preserve fine-grained regional details in image and video captions, while large vision-language models such as Qwen-based systems leverage large-scale multi-modal pretraining to produce semantically consistent and expressive descriptions across diverse visual content \cite{bai2023qwen,yang2025qwen3}.

In recent years, advances in latent diffusion models (LDMs) \cite{ho2020denoising,song2020denoising,rombach2022high} have led to significant progress in generative visual modeling, substantially improving the ability to synthesize realistic video content from sparse semantic inputs. Image-to-video and text-to-video generation models \cite{NEURIPS2023_52f05049,ICCV2023} typically operate in a latent space using U-Net–based denoising architectures \cite{Bao_2023_CVPR}, while incorporating semantic conditioning through pretrained encoders such as CLIP for cross-modal image–text alignment or large-scale language models such as T5 for contextual text representation \cite{xu2024easyanimate}. By conditioning the diffusion process on these high-level semantic embeddings, such models can generate temporally coherent and visually plausible video sequences without requiring access to the original video signal. Through joint modeling of motion dynamics, temporal consistency, and high-level semantics, these generative models are able to infer and synthesize missing visual details, effectively “imagining” plausible content that satisfies the provided textual or visual constraints \cite{xu2024easyanimate,guo2023animatediff,chen2024videocrafter2}.

\section{Motivation}


Existing approaches to video communication struggle to reconcile the tension
between \emph{high-dimensional semantic representation} and \emph{efficient transmission}.
This motivates a fundamental question: \emph{can video communication be achieved
without transmitting raw visual data directly?}
We observe that video content can be decomposed into \textit{spatial information} and \textit{temporal information}, which enables compact semantic representations for reducing communication overhead.


\subsection{Rethinking Video Communication}

Existing wireless communication systems are designed to ensure reliable delivery
and accurate reconstruction of transmitted signals, which is reflected across
the communication stack from channel coding and retransmission mechanisms to
video streaming protocols that deliver compressed visual data for playback at
the receiver \cite{Wei10740493,ieee10217138,OUYANG2025125665}. In video
communication scenarios, this objective translates into transmitting visual
data itself, where communication cost is directly related to video resolution,
frame rate, and compression fidelity. Even when downstream tasks require only
task-relevant semantics rather than full visual detail, substantial bandwidth
remains necessary to support signal reconstruction. This reconstruction-oriented
pipeline also exposes raw visual information by default, which makes privacy
control and selective disclosure difficult to support within the communication
process \cite{10621341,fl_wu}. Moreover, uniform signal delivery prevents the
sender, the network, and the receiver from independently expressing constraints
on semantic representation, transmission efficiency, and content presentation.

Emerging communication scenarios increasingly challenge the assumption that
faithful signal reconstruction is either necessary or sufficient. In
next-generation wireless systems, communication tasks are often driven by
trust, decision-making, and perception, where success depends on whether the
receiver can correctly interpret semantic intent and perform subsequent
actions. For example, wireless blockchain networks for 6G emphasize traceability,
consistency, and reliable information exchange between distributed nodes,
which does not require raw signal transmission \cite{Luo11105407}. Intelligent
wireless signal recognition similarly focuses on discriminative semantic
features for classification, security, and spectrum awareness rather than
reconstructing original waveforms \cite{Zhang11002534}.

At the same time, immersive and industrial applications impose heterogeneous
constraints on communication systems. Wireless virtual and augmented reality
requires ultra-low latency, high reliability, and adaptive resource allocation
under strict bandwidth and power limitations \cite{Mohsin11175685}. Industrial
communication environments such as 6G, Wi-Fi8, and time-sensitive networking
prioritize deterministic latency, robustness, and controllability to support
safety-critical operations \cite{John2023Industry}. As privacy regulations
evolve and user awareness increases, many scenarios explicitly discourage
transmission of raw visual signals and instead require controllable abstraction
and selective disclosure. These trends reveal a fundamental mismatch between
reconstruction-centric video transmission and modern requirements for
communication-efficient semantic delivery, which motivates a visual data-free
communication framework such as \textit{Mirage}.

\subsection{Generative Reconstruction Opportunity}

Recent advances in AI-generated models fundamentally alter the role of signal reconstruction in communication systems. Modern generative models are capable of synthesizing high-quality visual content from sparse semantic representations, such as textual descriptions, key frames, or latent features, without requiring access to the original video signal \cite{rombach2022high, guo2023animatediff, chen2024videocrafter2}. Rather than recovering missing pixels, these models infer perceptually plausible content that satisfies semantic, temporal, and contextual constraints. As a result, meaningful visual experiences can be obtained without transmitting raw visual signals, enabling communication to focus on delivering compact semantic cues and generation constraints while delegating perceptual synthesis to the receiver.

This shift is not merely driven by progress in generative vision models, but is increasingly reflected in networking and system-level designs that treat AI models as core communication primitives. Recent work such as NetLLM demonstrates that large language models can serve as foundation models for diverse networking tasks, processing multimodal inputs and generating task-specific decisions with strong generalization, significantly reducing task-specific model design overhead \cite{WuDuo3672268}. Similarly, MixNet shows that the dynamic communication patterns induced by mixture-of-expert models fundamentally challenge static network designs, requiring adaptive, model-aware communication substrates \cite{Liao3750465}. In parallel, privacy-preserving transformer service frameworks such as SCX, illustrate how intermediate semantic states, rather than raw inputs or outputs, can become the primary objects of communication, with explicit consideration of latency, efficiency, and security \cite{Yuan3750509}. These systems indicate that AI-generated models are no longer passive endpoints but actively reshape communication requirements and architectures.

Consequently, generation must be treated as an integral component of the communication pipeline rather than an auxiliary application-layer function. Communication architectures must explicitly account for semantic representation, controllable abstraction, and receiver-side synthesis, shifting the design objective from faithful signal delivery to intent-aligned perceptual outcomes. This paradigm motivates the need for new semantic communication frameworks, such as Mirage, that tightly integrate representation learning, efficient transmission, and generative reconstruction under system-level constraints.

\section{Mirage System Architecture}

\begin{figure*}[!t]
  \centering \includegraphics[width=0.95\linewidth]{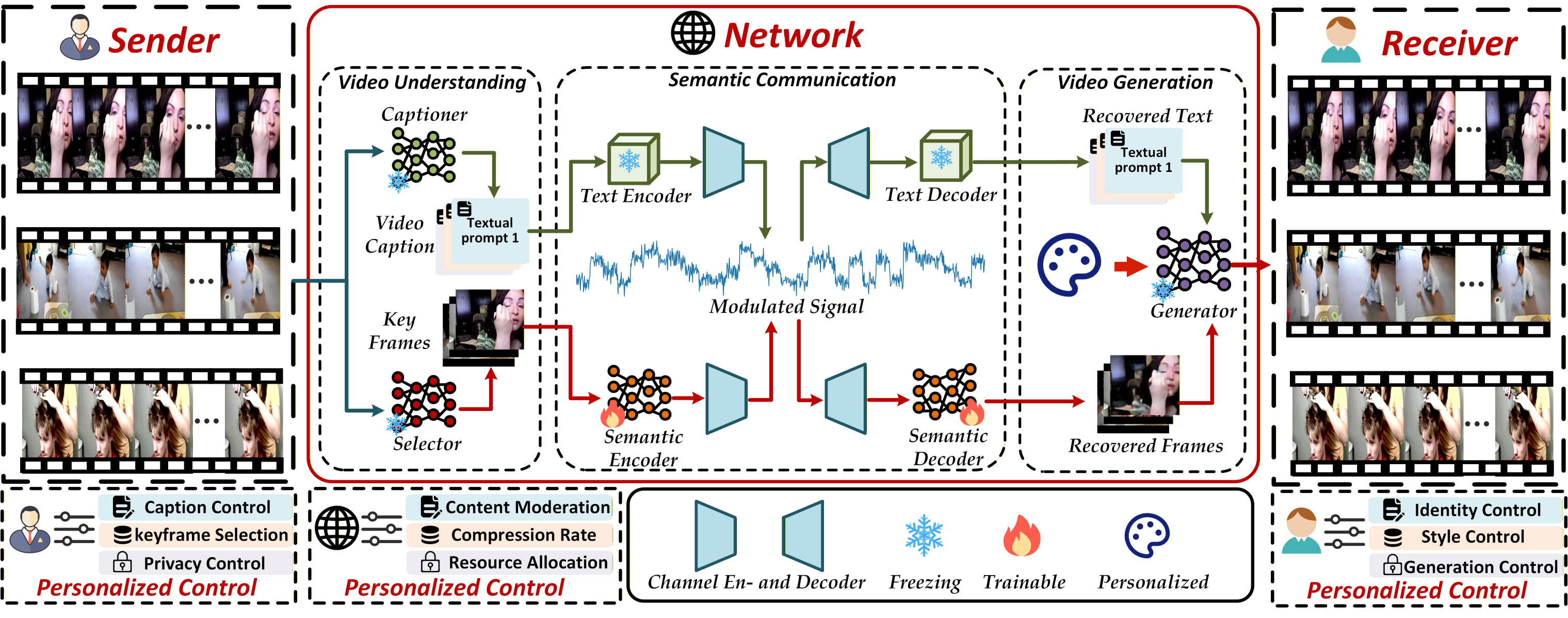}
 \caption{Overview of the Mirage architecture.
Mirage consists of three components: a sender-side video understanding module that converts raw video into semantic representations, a network-side semantic communication module for efficient transmission, and a receiver-side video generation module for personalized synthesis. The figure illustrates the roles of the sender, the network, and the receiver, where semantic representation, adaptive transmission, and receiver-side generation together enable communication-efficient and personalized video delivery without transmitting visual data.}
  \Description{}
  \label{fig:framework}
\end{figure*}

\subsection{Overview of the Mirage}

We provide a detailed overview of the design of Mirage, a novel framework for future communication paradigms that effectively addresses the challenges outlined in the previous section. As illustrated in Figure \ref{fig:framework},
\textit{Mirage} consists of three stages: video understanding, semantic communication,
and video generation.
On the sender side, a captioner extracts temporal dynamics through video captioning,
where the resulting text is further compressed losslessly by a text encoder.
Meanwhile, a selector identifies intent-aligned keyframes, and a semantic encoder
produces compact spatial representations to reduce spatial redundancy.
Together, these components form a low-dimensional multimodal representation of the video. At the receiver, Mirage performs generative reconstruction instead of traditional decoding.
Spatial representations serve as visual anchors, while temporal instructions act as
conditional guidance for the generator.
Through conditional denoising, the generator reconstructs a video that preserves
semantic consistency with the original content.
We refer to the transmitted spatial and temporal representations as
\emph{perceptual illusion}, which enables visual-data-free video communication
without transmitting raw video signals and provides improved privacy protection.

\subsection{Sender-Side Video Understanding}

\begin{figure}[!t]
  \centering
  \includegraphics[width=\linewidth]{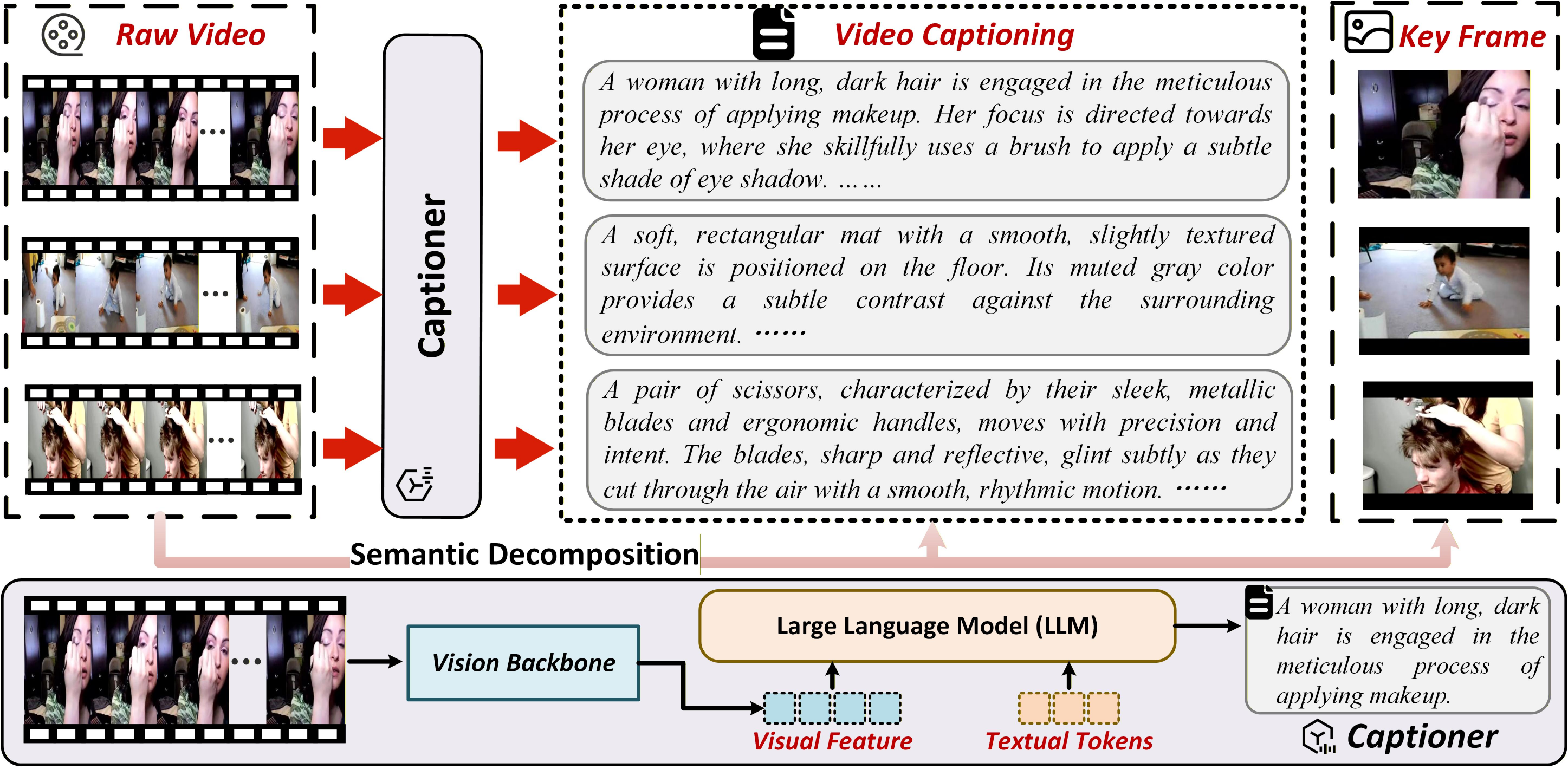}
  \caption{Sender-Side Video Understanding via Semantic Decomposition.}

  \Description{A woman and a girl in white dresses sit in an open car.}
  \label{fig:semantic_descomposition}
\end{figure}

On the sender side, Mirage decomposes video content into compact semantic representations suitable for communication, rather than encoding the video signal itself. The sender-side video understanding module separates video content into two complementary components: temporal sequence information captured by video captioning and spatial appearance information represented by selected key frames. Specifically, as shown in Figure \ref{fig:semantic_descomposition}, video understanding decouples the video content into two complementary modalities: \emph{video captioning} and \emph{key-frame selection}, forming a multimodal semantic information of the original video. Let $V = \{v_1, v_2, \ldots, v_n\}$ denote an input video sequence with $n$ frames, where each frame $v_i \in \mathbb{R}^{H \times W \times \mathcal{D}}$ represents an image with height $H$, width $W$, and $\mathcal{D}$ channels. The video understanding module on the sender-side aims to extract a semantic representation $S(V)$ from $V$, defined as:
\begin{equation}
S(V) = \big< C(V;\theta_{cap}),\; K(V; \theta_{sel}) \big>,
\end{equation}
where $C(V;\theta_{cap})$ denotes the semantic text description generated via video captioning, and $K(V; \theta_{sel})$ represents a set of selected key frames determined under the sender-side personalization parameters $\theta_{sel}$.


\paragraph{Video Captioning.}
To preserve temporal sequence information for communication, Mirage uses video captioning to encode motion dynamics and event-level semantics into textual descriptions. These captions serve as compact semantic carriers of temporal information rather than descriptive summaries of visual content. In our implementation, we adopt a DAM-based \cite{dam} captioning architecture as a part of the semantic encoder, which combines localized visual understanding with global contextual reasoning. A vision backbone with attention adapters is explicitly used to extract structured visual features from video frames, ensuring that semantic descriptions are grounded in visual evidence rather than purely language priors.

Given an input video $V=\{v_1,\dots,v_n\}$, each frame is first processed by the vision backbone to obtain a sequence of fused visual representations:
\begin{equation}
\{z'_1, z'_2, \dots, z'_n\} = \Psi_{\mathrm{vis}}(V;\theta_{vis}),
\end{equation}
where $\Psi_{\mathrm{vis}}(\cdot)$ denotes the vision backbone equipped with attention adapters that integrate global context and localized visual cues.

The visual features from all frames are concatenated along the temporal (sequence) dimension and fed into a large language model to generate a video-level semantic description:
\begin{equation}
C(V;\theta_{\mathrm{cap}}) = \mathrm{LLM}\big(t, \{z'_1, z'_2, \dots, z'_n\};\theta_{llm}\big),
\end{equation}
where $t$ denotes optional textual prompt tokens, $C(V;\theta_{\mathrm{cap}})$ is a sequence of semantic tokens describing the video content across frames and $\theta_{\mathrm{cap}}=\big< \theta_{vis},\theta_{llm}\big>$.

Unlike conventional video captioning systems that aim for maximal descriptiveness, Mirage treats captions as \emph{semantic carriers}. Their content, granularity, and level of abstraction can be adjusted according to sender-side personalization objectives, such as privacy preservation, semantic emphasis, or generation controllability. As a result, the generated captions are optimized to support communication-efficient transmission and receiver-side generative video synthesis, rather than faithful visual reconstruction.

\paragraph{Key-Frame Selection.}

In parallel, video understanding selects a subset of representative frames from the original video to serve as visual semantic anchors for downstream generation. Given sender-side preferences $\theta_{{sel}}$ (e.g., privacy level, visual fidelity, or bandwidth budget), each frame is first assigned a semantic relevance score conditioned on personalization requirements. Therefore, keyframe selection can be formulated as follows:

\begin{equation}
K(V;\theta_{{sel}}) =
\operatorname*{arg\,max}_{\substack{\mathcal{S} \subseteq \{1,\dots,n\} \\    |\mathcal{S}| = N}}
\sum_{t \in \mathcal{S}} \mathrm{Score}(\mathbf{m}_t, \mathbf{u}_S;\theta_{sel}),
\end{equation}
where $\mathcal{S}$ represents a candidate subset of frame indices with fixed cardinality $|\mathcal{S}|=N$. $\mathbf{m}_t$ denotes the semantic representation extracted from frame $v_t$, while $\mathbf{u}_S$ captures sender-side preferences or task requirements. The scoring function $\mathrm{Score}(\mathbf{m}_t,\mathbf{u}_S;\theta_{\mathrm{sel}})$ assigns each frame a semantic importance value conditioned on both its content and sender-side objectives, and the operator $\arg\max$ selects the subset of frames that maximizes the general semantic utility under the given budget. In practice, the scoring function can be directly determined by the sender-side or implemented using existing vision language models such as CLIP \cite{radford2021clip} or OpenCLIP \cite{Cherti_2023_CVPR} to measure semantic relevance, optionally combined with privacy-aware detectors (e.g., YOLO \cite{lei2025yolov13}) depending on application requirements.



\paragraph{Personalized Semantic Representation.}
In Mirage, sender-side personalization is introduced through the parameterized captioning function $C(V;\theta_{{cap}})$ and the key-frame selection function $K(V;\theta_{{sel}})$. By adjusting $\theta_{{cap}}$, the sender controls the granularity and semantic exposure of caption tokens, while $\theta_{{sel}}$ determines which frames are selected as visual semantic anchors under privacy, fidelity, or bandwidth constraints.  This sender-side control constitutes the semantic representation component of Mirage personalization, which complements network-side transmission control and receiver-side generative rendering. As a result, the output of video understanding forms a flexible multimodal semantic representation that supports communication-efficient and personalized video delivery without relying on pixel-level encoding.



\subsection{Semantic Communication}
\label{sec:semantic-communication}

\begin{figure}[!t]
  \centering
  \includegraphics[width=\linewidth]{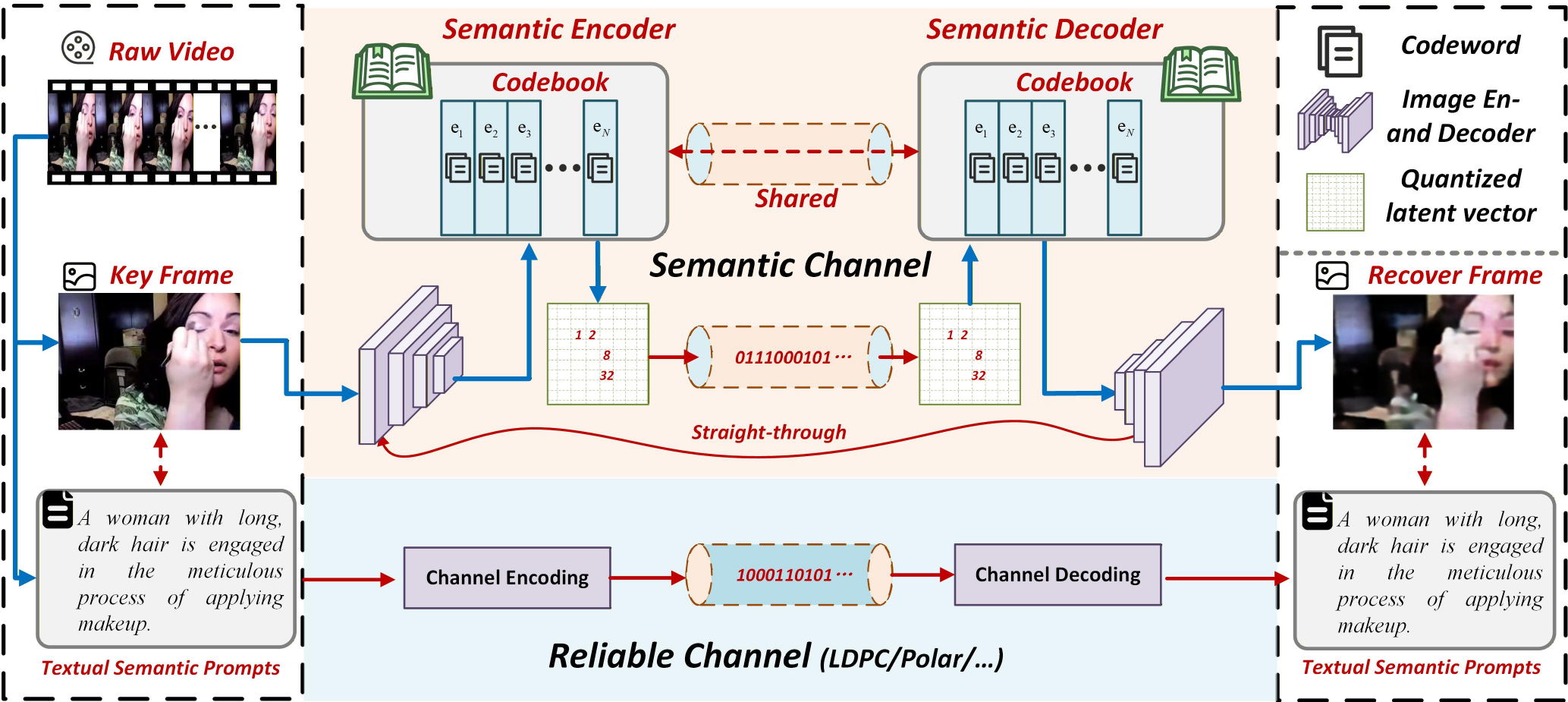}
  \caption{Semantic communication in Mirage.
Instead of transmitting visual data, the sender converts video content into compact semantic representations consisting of textual descriptions and key-frame representations.}
  \Description{A woman and a girl in white dresses sit in an open car.}
  \label{fig:video_semantic1}
\end{figure}

Mirage reformulates video communication as the transmission of compact semantic payloads rather than pixel-level signals.
The objective is not to reconstruct the original video signal, but to preserve semantic and perceptual consistency under strict communication constraints.
To this end, as shown in Figure \ref{fig:video_semantic1}, Mirage decomposes video content into two complementary semantic streams: 
(i) visually grounded keyframe semantics transmitted at ultra-low bitrate, and 
(ii) textual prompts transmitted with high reliability due to their small size and semantic sensitivity.
\paragraph{Wireless Channel Model}
To analyze how wireless channel variations affect visual reconstruction quality, we consider an additive white Gaussian noise (AWGN) channel model \cite{guo2025,fl_wu}, in which the received signal is given by:
\begin{equation}
\hat S(V) = S(V) + \mathbf{n},
\end{equation}
where $\mathbf{n}$ consists of independent and identically distributed samples following $\mathcal{CN}(\mathbf{0}, \sigma_\mathbf{n}^2 \mathbf{I})$, and $\sigma_\mathbf{n}^2$ denotes the average noise power. Under the AWGN channel, the received signal-to-noise ratio (SNR) is defined as:
\begin{equation}
\mathrm{SNR}
= \frac{\mathbb{E}\!\left[\|S(V)\|^2\right]}{\mathbb{E}\!\left[\|\mathbf{n}\|^2\right]}
= \frac{P_S}{\sigma_\mathbf{n}^2},
\end{equation}
where $P_S = \mathbb{E}[\|S(V)\|^2]$ denotes the average power of the transmitted signal. Given the resulting SNR, we further characterize the maximum achievable transmission capacity using the Shannon formula. Specifically, for the receiver, the maximum throughput can
be formulated as follows:
\begin{equation}
\mathcal{T} = b \log_2\!\left(1 + \mathrm{SNR}\right),
\end{equation}
where $b$ denotes the allocated bandwidth and $\mathcal{T}$ denotes the maximum transmission throughput.

While SNR characterizes the average channel quality, it does not directly reflect symbol-level errors caused by channel noise. Since such errors can propagate through the decoding process and affect visual reconstruction, we further model the wireless channel using the bit error rate (BER). To compute the BER under coherent BPSK transmission, the SNR is first mapped to the bit-energy-to-noise density ratio $E_b/N_0$ as follows:
\begin{equation}
\frac{E_b}{N_0}
=
\frac{\mathrm{SNR}}{\eta},
\end{equation}
where $\eta = \mathcal{T}/b$ denotes the spectral efficiency in bits/s/Hz. Assuming ideal channel coding and transmission through capacity, we approximate $\eta \approx \log_2(1+\mathrm{SNR})$. The resulting BER can be expressed as follows:
\begin{equation}
\mathrm{BER}
=
\frac{1}{2}\,\mathrm{erfc}
\!\left(
\sqrt{
\frac{\mathrm{SNR}}{\log_2(1+\mathrm{SNR})}
}
\right),
\end{equation}
where $\mathrm{erfc}(\cdot)$ denotes the complementary error function, defined as
$\mathrm{erfc}(x) = \frac{2}{\sqrt{\pi}} \int_{x}^{\infty} e^{-t^2}\, dt.$

\paragraph{Discrete Semantic Encoding and Decoding.}
For each selected key-frame $\mathbf{x}_{i} \in K(V)$, Mirage applies a discrete semantic encoder for an extremely low compression rate.
The encoder produces a latent feature as follows:
\begin{equation}
\mathbf{z}_e = E_\phi(\mathbf{x}_{i}),
\end{equation}
where $E_\phi$ is the encoder, $c,h,w$ denotes the latent size of the channel, the height, and the width, respectively, and $\mathbf{z}_e \in \mathbb{R}^{c \times h \times w}$. A shared vector quantization codebook can be expressed as follows:
\begin{equation}
\mathcal{E} = \{ \mathbf{e}_1,\mathbf{e}_2\ldots, \mathbf{e}_K\},
\end{equation}
where $\mathbf{e}_k$ is the $k$-th codeword, $K$ is the size of the codebook.
Each latent vector is mapped to its nearest codeword in the codebook via vector quantization:
\begin{equation}
s^{j} = \operatorname*{arg\,max}_{k \in \{1,\dots,K\}} 
\left\| \mathbf{z}_e^{j} - \mathbf{e}_k \right\|_2^2,
\end{equation}
where $\mathbf{z}_e^{j} \in \mathbb{R}^d$ denotes the $j$-th latent vector of $\mathbf{z}_e$ produced by the encoder, and $\|\cdot\|_2$ denotes the Euclidean norm. As a result, each keyframe is represented by a tensor of discrete codeword indices $\mathbf{s} = \{ s^{j} \}_{j=1}^{h\times w}$, where $h$ and $w$ denote the spatial resolution of the latent feature map, and each index $s^{j} \in \{1,\dots,K\}$ specifies the selected codeword for the corresponding spatial location. Therefore, the communication cost for transmitting a quantized keyframe is approximately $B_{\mathrm{VQ}} \approx h\,w\,\log_2 K $ \text{bits}, assuming a fixed-length encoding for each codeword index.

At the receiver, channel decoding recovers $\hat{\mathbf{s}}=\{ \hat{s}^{j} \}_{j=1}^{h\times w}$, and perceptual reconstruction is obtained through the following:
\begin{equation}
\hat{\mathbf{x}}_{i} = D_\psi(\mathbf{e}_{\hat{s}}),
\end{equation}
where $D_\psi(\cdot)$ denotes the decoder. 

In addition, to ensure the semantic fidelity of the transmission of latent representation, we optimize the encoder and decoder using the following loss \cite{oord2017vqvae}:
\begin{equation}
\mathcal{L}_{\text{sem}}
=
\log p\!\left( \mathbf{x}_{i} \mid \mathbf{e}_{\mathbf{s}} \right)
+
\left\| \operatorname{sg}\!\left[\mathbf{z}_e\right] - \mathbf{e}_{\mathbf{s}} \right\|_2^2
+
\beta \left\| \mathbf{z}_e - \operatorname{sg}\!\left[\mathbf{e}_{\mathbf{s}}\right] \right\|_2^2 .
\end{equation}
where the operator $\operatorname{sg}[\cdot]$ denotes the stop-gradient operation, and
the loss jointly encourages faithful semantic reconstruction and consistency between encoder outputs and selected codewords, with $\beta$ controlling the commitment strength.

\paragraph{High-Reliability Transmission of Video Captioning.}

Video captioning usually has a low data volume but plays a directly critical guiding role in the downstream video generation process at the receiver. To ensure accurate delivery of such control semantics, Mirage transmits captioning information using a high-reliability communication mode. Prior to transmission, the video captioning is first compressed using a lossless compression scheme (e.g. \texttt{zlib}) to reduce redundancy while preserving exact semantic content. The resulting compressed bitstream is then protected by reliable channel coding. Let $\mathbf{b} \in \{0,1\}$ denote the losslessly compressed prompt bitstream.Reliable transmission is ensured by enforcing the following condition:
\begin{equation}
\Pr\!\left( \hat{\mathbf{b}} \neq \mathbf{b} \right) \le \epsilon, 
\quad \epsilon \ll 1 ,
\end{equation}
which guaranties near-error-free delivery of caption semantics for subsequent semantic decoding and video generation. where $\mathbf{b} \in \{0,1\}$ denotes the transmitted bitstream corresponding to the compressed textual prompt,
$\hat{\mathbf{b}}$ represents its decoded version at the receiver,
and $B_P$ is the number of bits after lossless compression.
The probability $\Pr(\hat{\mathbf{b}} \neq \mathbf{b})$ characterizes the block error probability, i.e., the probability that at least one bit in the prompt is incorrectly decoded.
The parameter $\epsilon$ specifies the target reliability level, with $\epsilon \ll 1$ indicating a highly reliable transmission. Therefore, the video captioning $C(V)$ transmitted from the receiving end is encoded using reliable channel coding, and the video captioning received after channel decoding can be denoted as $\hat{C}(V)$.

\paragraph{Personalized Network.}
Beyond sender- and receiver-side personalization, Mirage also supports network-side personalization within the semantic communication layer. The network can adapt the semantic encoder ${E}_{\phi}$ to channel conditions, such as instantaneous SNR, by adjusting the encoder’s downsampling ratio or latent dimensionality, thereby tuning the compression rate to balance robustness and efficiency. The network may further enforce privacy or policy constraints by filtering sensitive semantic content, for example by dropping selected key frames or modifying caption tokens, without reconstructing raw visual signals. At the communication level, Mirage is compatible with intelligent network control mechanisms, where learning-based or deep reinforcement learning (DRL)-driven controllers~\cite{Wudm1,Yindrl} jointly optimize bandwidth, transmission power, and beamforming based on channel states, task requirements, and semantic importance. This enables adaptive and task-aware resource management while preserving transmitted semantic information.

\subsection{Receiver-Side Personalized Generation}
\label{sec:receiver-generation}

\begin{figure}[!t]
  \centering
  \includegraphics[width=\linewidth]{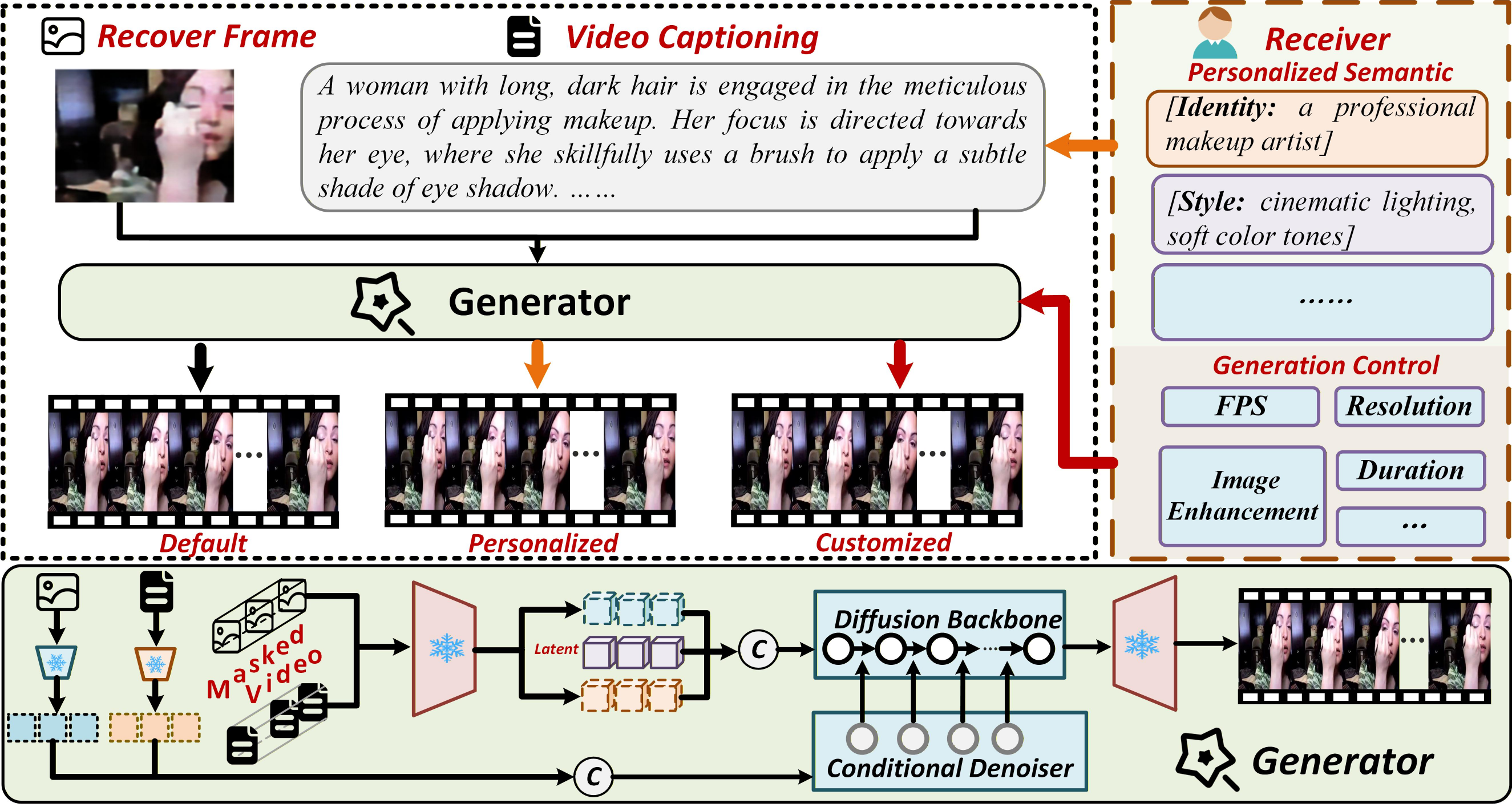}
  \caption{Receiver-side personalized generative reconstruction in Mirage.
Upon receiving semantic payloads, including decoded keyframe semantics and textual prompts, the receiver performs semantic-conditioned video generation instead of signal-level decoding.
Reconstructed keyframes are optionally augmented to improve robustness and diversity, while textual prompts can be adapted according to receiver-side preferences.}
  \Description{XXX.}
  \label{fig:generation}
\end{figure}


Upon receiving the semantic payloads transmitted by Mirage, the receiver reconstructs the video content through a personalized generative pipeline, as shown in Figure~\ref{fig:generation}.
Instead of reproducing the original signal, Mirage performs \emph{semantic-conditioned generation} driven by
$\hat{S}(V)=\langle \hat{C}(V),\,\hat{K}(V)\rangle$, where
$\hat{K}(V)=\{\hat{\mathbf{x}}_{i}\}_{i=1}^{\mathcal{R}}$ denotes the decoded keyframe semantics and
$\hat{C}(V)=\{y_1,y_2,\ldots,y_m\}$ denotes the decoded video captioning tokens.

\paragraph{Receiver Personalization Configuration.}
Mirage exposes a receiver-side personalization policy that (i) rewrites caption semantics for controllable generation and (ii) configures the video generator to match local preferences and system constraints.

\emph{(a) Caption personalization.}
Starting from $\hat{C}(V)$, the receiver constructs an effective generation prompt by injecting explicit \texttt{content}/\texttt{identity}/\texttt{style} fields:
\begin{equation}
\tilde{C}(V)=\mathcal{G}_{\text{cap}}\!\left(\hat{C}(V),\,\mathbf{u}_R\right)
=
\Big[\texttt{content}\ \Vert\ \texttt{identity}\ \Vert\ \texttt{style}\Big],
\label{eq:prompt_personalize}
\end{equation}
where $\mathcal{G}_{\text{cap}}(\cdot)$ denotes a lightweight transformation that converts captions into personalized prompts, and $\mathbf{u}_R$ encodes receiver-side preferences such as privacy constraints, target generation style, and safety requirements. The \texttt{identity} field governs subject identity attributes and supports operations such as anonymization or role substitution to enforce privacy, while the \texttt{style} field specifies the desired visual appearance, including the rendering style, lighting conditions, and color characteristics.

\emph{(b) Generator control.}
The receiver also selects a generator configuration vector as follows:
\begin{equation}
\boldsymbol{\theta}_g=\big\{F,\ \rho,\ (H_g,W_g),\ \lambda,\ N,\ \omega\big\},
\label{eq:gen_ctrl}
\end{equation}
where $F$ is the number of output frames, $\rho$ is the target frame rate (FPS), $(H_g,W_g)$ is the output resolution, $\lambda$ controls the strength of keyframe conditioning, $N$ is the number of denoising steps, and $\omega$ is the classifier-free guidance scale or an analogous guidance hyperparameter.

\paragraph{Personalized Semantic-Conditioned Video Generation.}
Given the personalized prompt $\tilde{C}(V)$ and the keyframe set
$\hat{K}(V)=\{\hat{\mathbf{x}}_{i}\}_{i=1}^{\mathcal{R}}$,
the receiver generates the output video using a conditional generator $p_{\psi}$:
\begin{equation}
\hat{V}\sim p_{\psi}\!\left(V \mid \hat{K}(V),\ \tilde{C}(V);\boldsymbol{\theta}_g\right),
\label{eq:cond_video_gen}
\end{equation}
where $\boldsymbol{\theta}_g$ specifies receiver-side generation controls, enabling the same transmitted semantics to produce multiple personalized video realizations.

In Mirage, $p_{\psi}$ is instantiated as a diffusion-based video generator operating in a latent space \cite{guo2023animatediff,xu2024easyanimate}.
Let $V=\{v_1,\ldots,v_F\}$ denote a video clip with $F$ frames.
Each frame is encoded by a pre-trained video auto-encoder into latent variables
$Z_0=\{z_0^1,\ldots,z_0^F\}$.
During inference, the diffusion backbone performs conditional denoising on the latent variables, progressively eliminating Gaussian noise under semantic and visual conditioning. The process can be formalized as follows:
\begin{equation}
Z_{t-1}
=
\mathcal{D}_{\psi}\!\left(
Z_t,\ t,\ \Phi_{\text{text}}(\tilde{C}(V)),\ \Phi_{\text{img}}(\hat{K}(V));\boldsymbol{\theta}_g
\right),
\label{eq:video_diffusion}
\end{equation}
where $t=T,\ldots,1$, $\mathcal{D}_{\psi}$ denotes the video denoising network with temporal modeling, 
and $\Phi_{\text{text}}(\cdot)$ and $\Phi_{\text{img}}(\cdot)$ encode textual and visual conditions, respectively.
The final video is obtained by decoding the denoised latent sequence as $\hat{V}=D_{\text{vae}}(Z_0)$.


\begin{figure*}[ht!]
    \centering
    \begin{subfigure}[t]{0.33\linewidth}
        \centering
        \includegraphics[width=\linewidth]{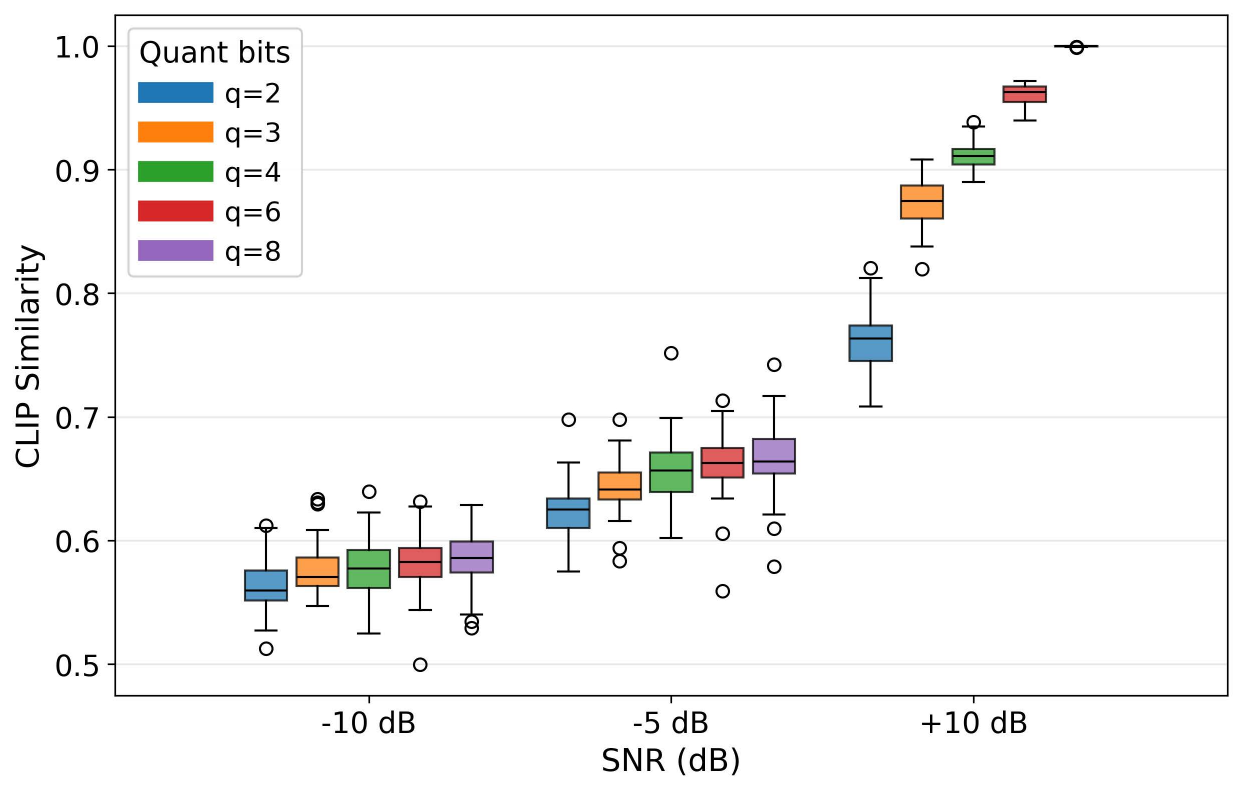}
        \caption{Semantic Similarity (CLIP) (↑ better)}
    \end{subfigure}
    \begin{subfigure}[t]{0.33\linewidth}
        \centering
        \includegraphics[width=\linewidth]{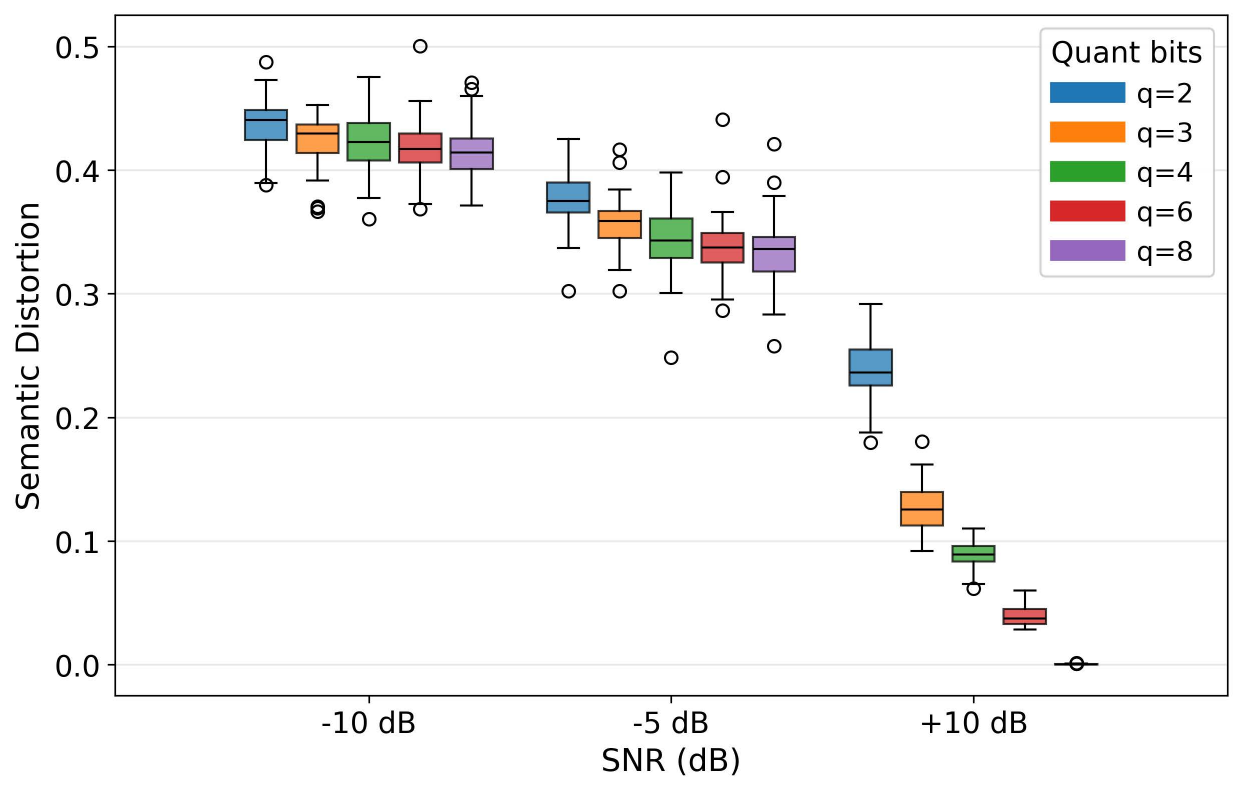}
        \caption{Semantic Distortion (↓ better)}
    \end{subfigure}
    \begin{subfigure}[t]{0.33\linewidth}
        \centering
        \includegraphics[width=\linewidth]{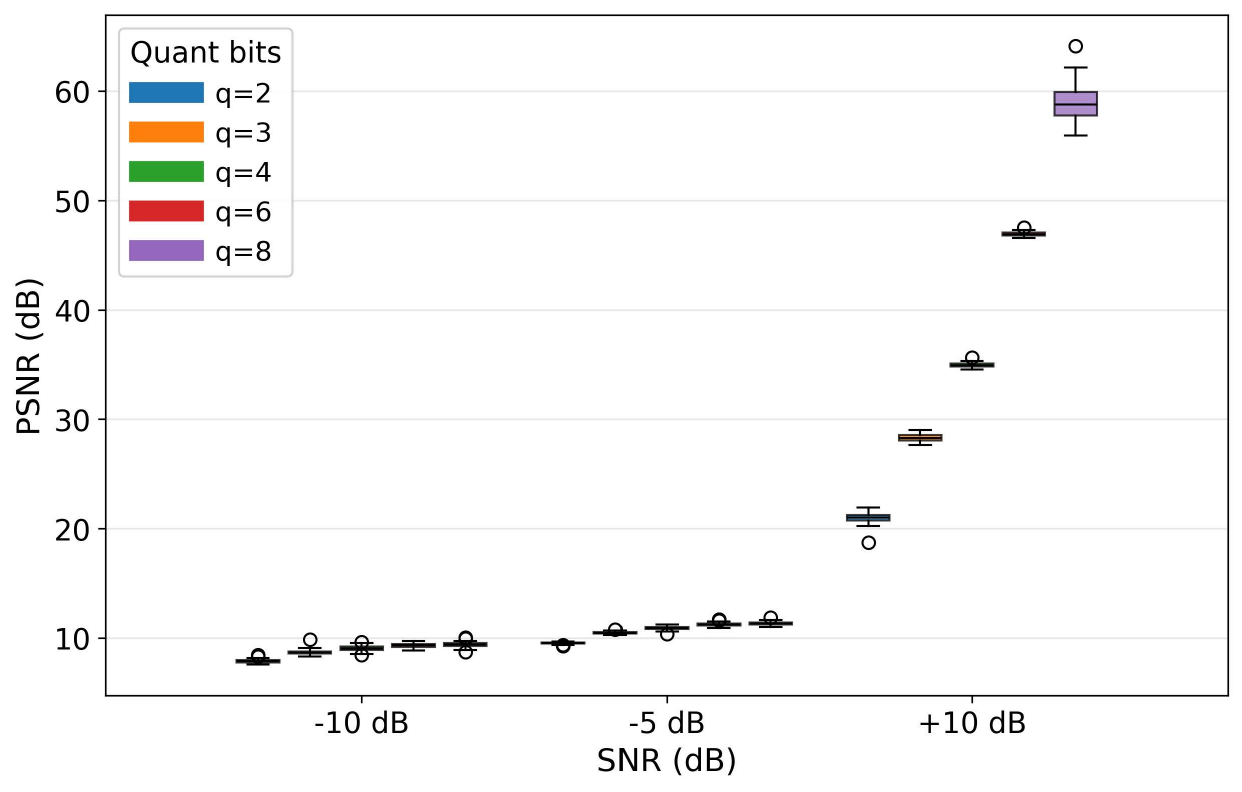}
        \caption{Reconstruction Quality (PSNR) (↑ better)}
    \end{subfigure}
    \begin{subfigure}[t]{0.33\linewidth}
        \centering
        \includegraphics[width=\linewidth]{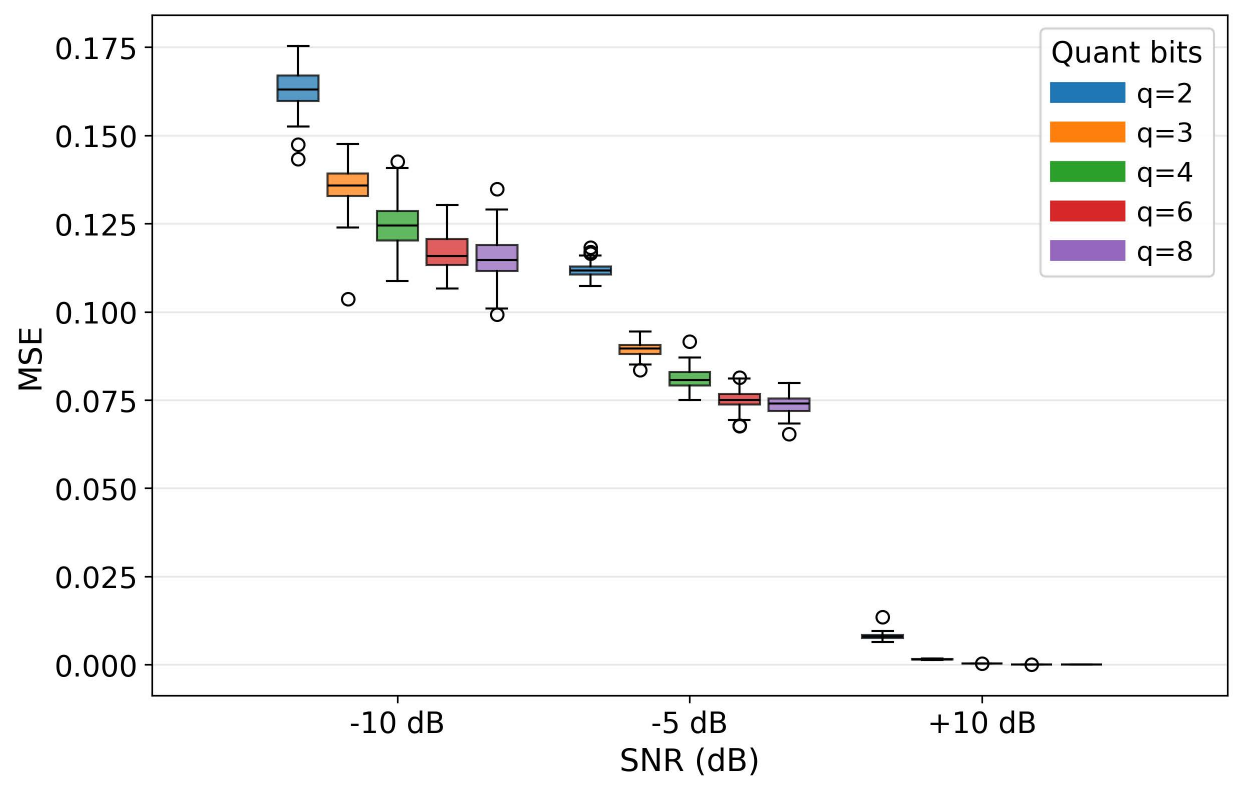}
        \caption{Pixel-Level Reconstruction Error (MSE)(↓ better)}
    \end{subfigure}
    \begin{subfigure}[t]{0.33\linewidth}
        \centering
        \includegraphics[width=\linewidth]{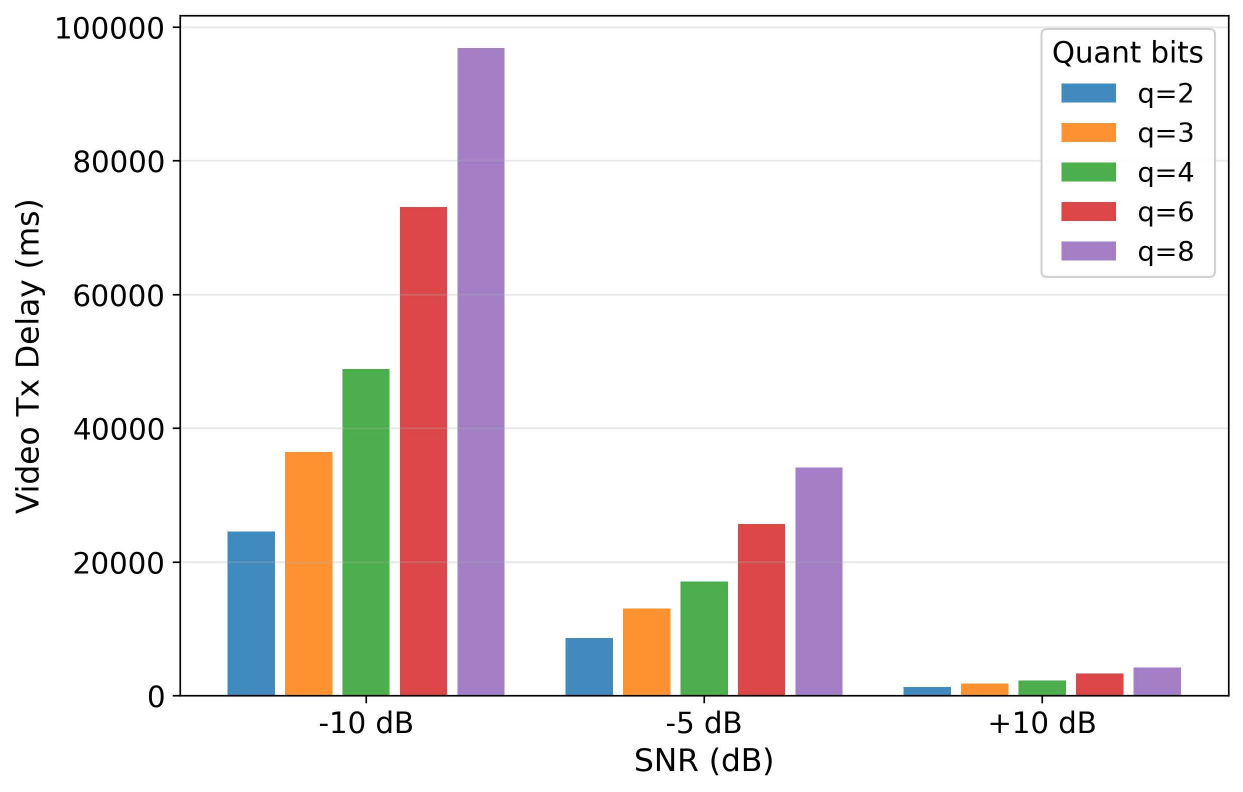}
        \caption{End-to-End Video Transmission Delay}
    \end{subfigure}
    \begin{subfigure}[t]{0.30\linewidth}
        \centering
        \includegraphics[width=\linewidth]{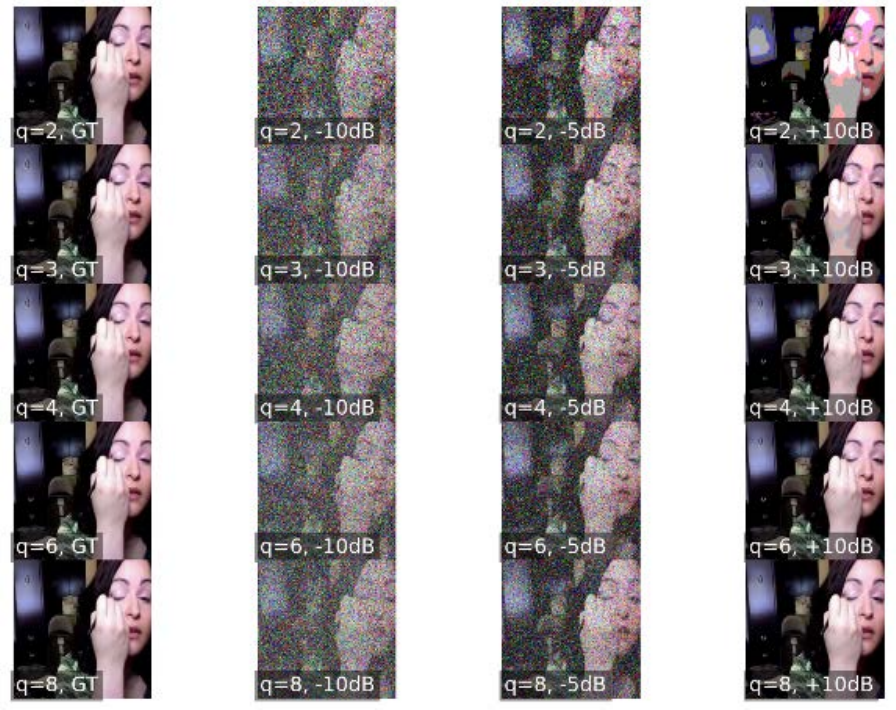}
        \caption{Visual Examples of Recovered Frames}
    \end{subfigure}

    \caption{Raw Video Transmission under Varying Wireless Channel Conditions. \emph{Box plots summarize the distribution across video samples, where circles indicate outliers corresponding to rare channel realizations or severe reconstruction failures}.}
    \label{fig:franklin-4sub}
\end{figure*}

\section{Evaluation}
\subsection{Experimental Setup}

\paragraph{Simulation Setup.}
We implement a pipeline of the proposed Mirage framework that integrates sender-side semantic feature encoding, communication-efficient semantic transmission, and receiver-side personalized video generation. The framework is experimentally tested on the publicly available video dataset UCF101 \cite{ucf101}. For video captioning in video understanding, Mirage adopts the NVIDIA Describe Anything Model (DAM) model \cite{dam} to extract high-level textual semantic descriptions and select representative key frames from input videos. These captions and key frames jointly serve as the multimodal semantic abstraction of the original video. For semantic communication, we construct a representation for efficient communication based on AE and VQ-VAE architectures, where semantic features are indexed by quantized continuous features and smaller-level discrete codes, respectively, for transmission over bandwidth-limited channels. The quantized level and size of the codebook can be adjusted to different wireless network conditions. On the receiver side, we employ state-of-the-art video generation models, including EasyAnimate \cite{xu2024easyanimate} and Grok Imagine \cite{GrokImagineAPI2024}, to synthesize personalized video content conditioned on recovered semantic representations and textual prompts. Furthermore, these components form an end-to-end Mirage framework that replaces pixel-level video transmission with semantic abstraction and perceptual generation. We evaluated Mirage in multiple simulated network environments with varying SNR, transmission quantization level, and BER, covering good, medium, and constrained wireless conditions. These settings are designed to capture the key factors that affect semantic transmission and generation quality in practical wireless networks.

\begin{table*}[ht!]
  \caption{Latency and Data Size Comparison of Different Transmission Schemes}
  \label{tab:latency_bitrate_horizontal}
  \begin{adjustbox}{width=0.95\textwidth}
  \begin{tabular}{lcccc}
    \toprule
    Scheme 
    & Raw (8qbit)
    & Conventional Semantic(AE-8qbit)
    & Mirage (AE-8qbit) 
    & Mirage (VQVAE-256) \\
    \midrule
    Transmission Mode
    & Raw streaming
    & Semantic compression
    & Key-frame + text
    & Key-frame + text \\
    Transmitted Content
    & Raw video frames
    & Semantic video features
    & AE key frames + text
    & VQVAE key frames + text \\
    Transmission data size (KB)  
    & 32415.88  
    & 21615.02    &128(frame)+0.40(text) 
    & 0.25(frame)+0.38(text)  \\
    Transmission latency (ms)
    & 96899.14
    & 64725.25 
    & 384.48 
    & 3.86 \\
    Bit per pixel (bpp)
    & 24.0 
    & 16.0 
    & 16.0 
    & 0.031 \\
    Data-Speedup
    & 1.0X 
    & 1.5X
    & 252.5X
    & \textbf{51818.7}X \\
    Latency-Speedup
    & 1.0X 
    & 1.5X
    & 252.0X 
    & \textbf{25081.0}X \\
    \bottomrule
  \end{tabular}
  \end{adjustbox}
\end{table*}

\paragraph{Metrics.}
We evaluate system performance from the perspectives of communication efficiency, perceptual quality, semantic consistency, and personalization. Communication efficiency is measured by transmission delay and bits per pixel. Perceptual quality is evaluated using perceptual and semantic metrics, including CLIP similarity and semantic distortion \cite{radford2021clip, wang2025reclip}, while peak signal-to-noise ratio (PSNR) and mean squared error (MSE) \cite{semantic1} are reported to reflect pixel-level differences. For personalized generation, we examine receiver-side video generation under different preference settings (e.g., style control and privacy adjustment) using recovered semantic representations from both AE- and VQ-VAE-based transmission. Finally, semantic consistency between captions generated from synthesized videos and those derived from raw videos is evaluated using BERTScore \cite{zhang2020bertscore} and Sentence-BERT cosine similarity \cite{reimers2019sentence}, capturing semantic alignment at the raw videos and generated videos.


\subsection{Communication Efficiency}



Table~\ref{tab:latency_bitrate_horizontal} summarizes the communication efficiency of different video transmission schemes in terms of transmitted data size, end-to-end latency, and effective bitrate. As a reconstruction-centric baseline, raw video streaming incurs extremely high communication overhead, requiring 32.4MB of transmitted data and nearly 97seconds of end-to-end latency under poor channel conditions of the evaluated setting. Even with conventional semantic compression based on AE features, the transmitted data volume remains above 21~MB, resulting in only marginal improvements in both data size and latency. In contrast, Mirage fundamentally changes the communication cost structure by transmitting only compact semantic representations. When using AE-based key frames and textual descriptions, Mirage reduces the transmitted data size to approximately 128~KB for visual content and 0.40~KB for text, yielding a data reduction of over two orders of magnitude compared to raw streaming. This reduction directly translates into a 252.5$\times$ data-speedup and a comparable 252.0$\times$ latency-speedup.

The efficiency gains become even more pronounced when adopting VQ-VAE-based semantic transmission. By discretizing semantic representations into compact code indices, Mirage further reduces the transmitted visual data to only 0.25~KB per key frame, with an additional 0.38~KB for text. As a result, the overall bit-per-pixel (bpp) drops from 24.0 in raw streaming to 0.031, representing a reduction of nearly three orders of magnitude. Correspondingly, Mirage with VQ-VAE achieves a data-speedup of \textbf{51,818.7}$\times$ and a latency-speedup of \textbf{25,081.0}$\times$ relative to raw transmission.

\begin{table}[!t]
\centering
\caption{Speedup Results at SNR $=-10$ dB.}
\label{tab:speedup_snr_m10}
\renewcommand{\arraystretch}{0.35}
\begin{adjustbox}{width=0.45\textwidth}
\begin{tabular}{>{\centering\arraybackslash}m{2cm} l cc}
\toprule
Quantization $q$ / Codebook $K$ & Scheme & Data-Speedup ($\times$) & Latency-Speedup ($\times$) \\
\midrule
\multirow{3}{*}{\centering 2}  & RAW      & 1.0 & 1.0 \\
                              & RAW-AE   & 1.5 & 1.5 \\
                              & Mirage-AE & 251.1 & 249.5 \\
\midrule
\multirow{3}{*}{\centering 3}  & RAW      & 1.0 & 1.0 \\
                              & RAW-AE   & 1.5 & 1.5 \\
                              & Mirage-AE & 250.5 & 249.4 \\
\midrule
\multirow{3}{*}{\centering 4}  & RAW      & 1.0 & 1.0 \\
                              & RAW-AE   & 1.5 & 1.5 \\
                              & Mirage-AE & 252.8 & 251.9 \\
\midrule
\multirow{3}{*}{\centering 6}  & RAW      & 1.0 & 1.0 \\
                              & RAW-AE   & 1.5 & 1.5 \\
                              & Mirage-AE & 253.2 & 252.6 \\
\midrule
\multirow{3}{*}{\centering 8}  & RAW      & 1.0 & 1.0 \\
                              & RAW-AE   & 1.5 & 1.5 \\
                              & Mirage-AE & 252.5 & 252.0 \\
\midrule
\centering 256  & Mirage-VQVAE & 51818.7 & 25081.0 \\
\centering 512  & Mirage-VQVAE & 49353.3 & 24490.9 \\
\centering 1024 & Mirage-VQVAE & 47111.8 & 23927.9 \\
\centering 2048 & Mirage-VQVAE & 45065.0 & 23390.3 \\
\bottomrule
\end{tabular}
\end{adjustbox}
\end{table}

\begin{table}[!t]
\centering
\caption{Speedup Results at SNR $=-5$ dB.}
\label{tab:speedup_snr_m5}
\renewcommand{\arraystretch}{0.35}
\begin{adjustbox}{width=0.45\textwidth}
\begin{tabular}{>{\centering\arraybackslash}m{2cm} l cc}
\toprule
Quantization $q$ / Codebook $K$ & Scheme & Data-Speedup ($\times$) & Latency-Speedup ($\times$) \\
\midrule
\multirow{3}{*}{\centering 2}  & RAW      & 1.0 & 1.0 \\
                              & RAW-AE   & 1.5 & 1.4 \\
                              & Mirage-AE & 247.7 & 243.1 \\
\midrule
\multirow{3}{*}{\centering 3}  & RAW      & 1.0 & 1.0 \\
                              & RAW-AE   & 1.5 & 1.5 \\
                              & Mirage-AE & 253.4 & 250.2 \\
\midrule
\multirow{3}{*}{\centering 4}  & RAW      & 1.0 & 1.0 \\
                              & RAW-AE   & 1.5 & 1.5 \\
                              & Mirage-AE & 251.6 & 249.1 \\
\midrule
\multirow{3}{*}{\centering 6}  & RAW      & 1.0 & 1.0 \\
                              & RAW-AE   & 1.5 & 1.5 \\
                              & Mirage-AE & 253.9 & 252.3 \\
\midrule
\multirow{3}{*}{\centering 8}  & RAW      & 1.0 & 1.0 \\
                              & RAW-AE   & 1.5 & 1.5 \\
                              & Mirage-AE & 254.3 & 253.1 \\
\midrule
\centering 256  & Mirage-VQVAE & 52195.9 & 12877.4 \\
\centering 512  & Mirage-VQVAE & 49712.5 & 12722.2 \\
\centering 1024 & Mirage-VQVAE & 47454.7 & 12570.6 \\
\centering 2048 & Mirage-VQVAE & 45393.0 & 12422.7 \\
\bottomrule
\end{tabular}
\end{adjustbox}
\end{table}

\begin{table}[!t]
\centering
\caption{Speedup Results at SNR $=10$ dB.}
\label{tab:speedup_snr_10}
\renewcommand{\arraystretch}{0.35}
\begin{adjustbox}{width=0.45\textwidth}
\begin{tabular}{>{\centering\arraybackslash}m{2cm} l cc}
\toprule
Quantization $q$ / Codebook $K$ & Scheme & Data-Speedup ($\times$) & Latency-Speedup ($\times$) \\
\midrule
\multirow{3}{*}{\centering 2}  & RAW      & 1.0 & 1.0 \\
                              & RAW-AE   & 1.5 & 1.3 \\
                              & Mirage-AE & 251.6 & 223.6 \\
\midrule
\multirow{3}{*}{\centering 3}  & RAW      & 1.0 & 1.0 \\
                              & RAW-AE   & 1.5 & 1.4 \\
                              & Mirage-AE & 253.6 & 232.1 \\
\midrule
\multirow{3}{*}{\centering 4}  & RAW      & 1.0 & 1.0 \\
                              & RAW-AE   & 1.5 & 1.4 \\
                              & Mirage-AE & 254.7 & 237.3 \\
\midrule
\multirow{3}{*}{\centering 6}  & RAW      & 1.0 & 1.0 \\
                              & RAW-AE   & 1.5 & 1.4 \\
                              & Mirage-AE & 258.4 & 245.6 \\
\midrule
\multirow{3}{*}{\centering 8}  & RAW      & 1.0 & 1.0 \\
                              & RAW-AE   & 1.5 & 1.4 \\
                              & Mirage-AE & 254.2 & 244.4 \\
\midrule
\centering 256  & Mirage-VQVAE & 52176.4 & 2027.2 \\
\centering 512  & Mirage-VQVAE & 49693.9 & 2023.6 \\
\centering 1024 & Mirage-VQVAE & 47437.0 & 2020.0 \\
\centering 2048 & Mirage-VQVAE & 45376.1 & 2016.4 \\
\bottomrule
\end{tabular}
\end{adjustbox}
\end{table}

Tables~\ref{tab:speedup_snr_m10}--\ref{tab:speedup_snr_10} report the data and latency speedup achieved by different transmission schemes under varying channel conditions, characterized by SNR levels of $-10$~dB, $-5$~dB, and $10$~dB. These results provide additional insights into how channel quality influences the efficiency gains of semantic and generative transmission. Across all SNR regimes, raw video transmission serves as the baseline with unit speedup, while conventional semantic compression using AE (RAW-AE) consistently achieves only modest improvements of approximately $1.3$--$1.5\times$. This indicates that feature-level compression alone offers limited benefits in reducing end-to-end transmission cost, regardless of channel quality. In contrast, Mirage-AE consistently achieves around $250\times$ data and latency speedup across all tested SNRs and quantization settings. In particular, these speedup values remain relatively stable as SNR improves, suggesting that the efficiency gains of Mirage-AE are primarily driven by semantic abstraction rather than favorable channel conditions. Even under severely degraded channels ($-10$~dB), Mirage-AE maintains a speedup comparable to that observed at higher SNRs. Mirage-VQVAE further amplifies these gains by several orders of magnitude. Under all channel conditions, Mirage-VQVAE achieves data speedups exceeding $40,000$, with peak values above $50,000$ for smaller codebooks. Although latency speedup decreases as SNR increases, due to higher achievable throughput under good channel conditions, Mirage-VQVAE still maintains latency improvements of over three orders of magnitude compared to raw transmission.




\begin{figure*}[ht!]
    \centering
    \begin{subfigure}[t]{0.33\linewidth}
        \centering
        \includegraphics[width=\linewidth]{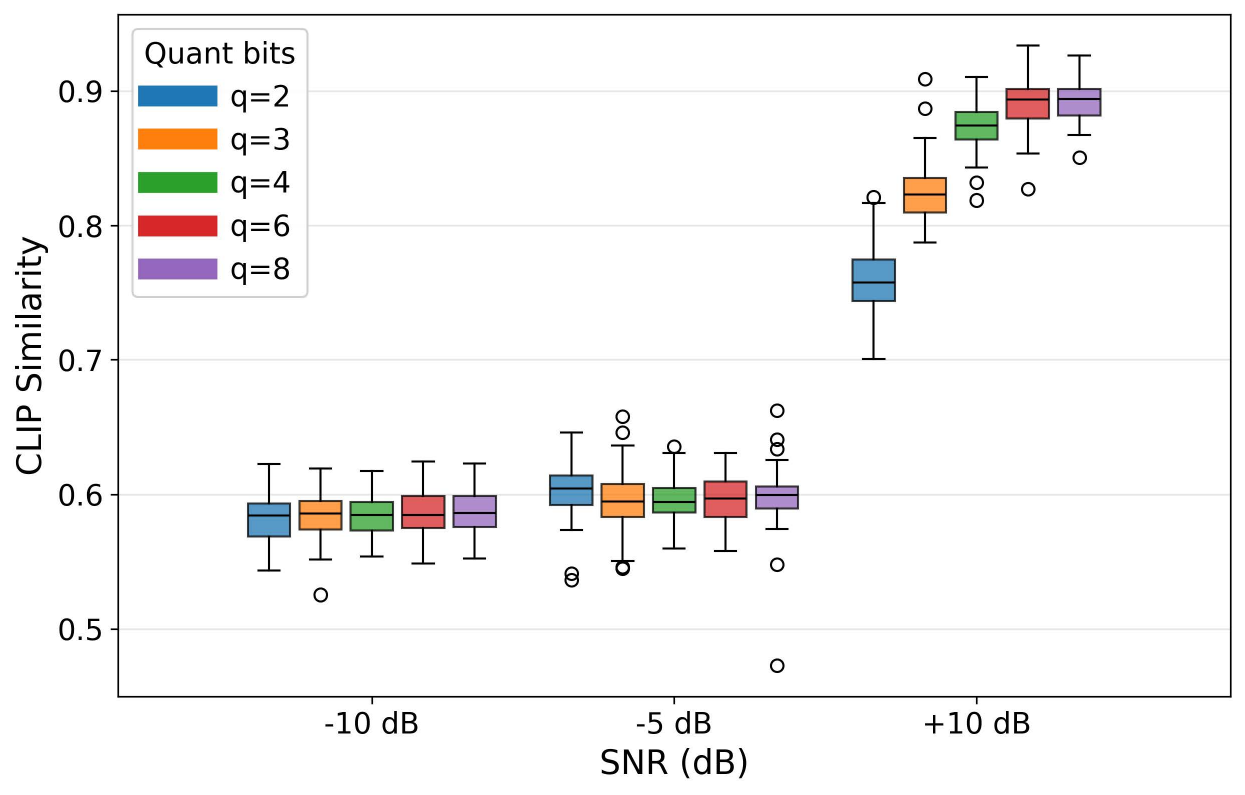}
        \caption{Semantic Similarity (CLIP) (↑ better)}
    \end{subfigure}
    \begin{subfigure}[t]{0.33\linewidth}
        \centering
        \includegraphics[width=\linewidth]{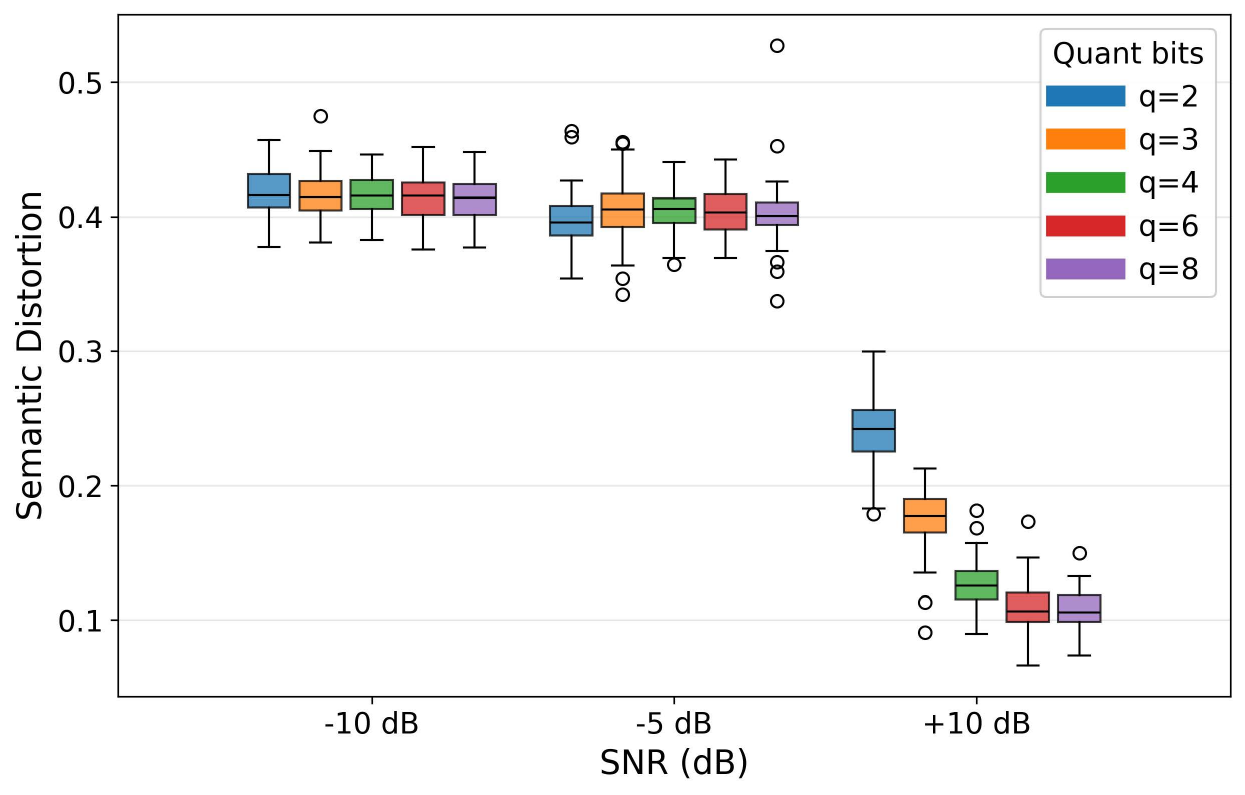}
        \caption{Semantic Distortion (↓ better)}
    \end{subfigure}
    \begin{subfigure}[t]{0.33\linewidth}
        \centering
        \includegraphics[width=\linewidth]{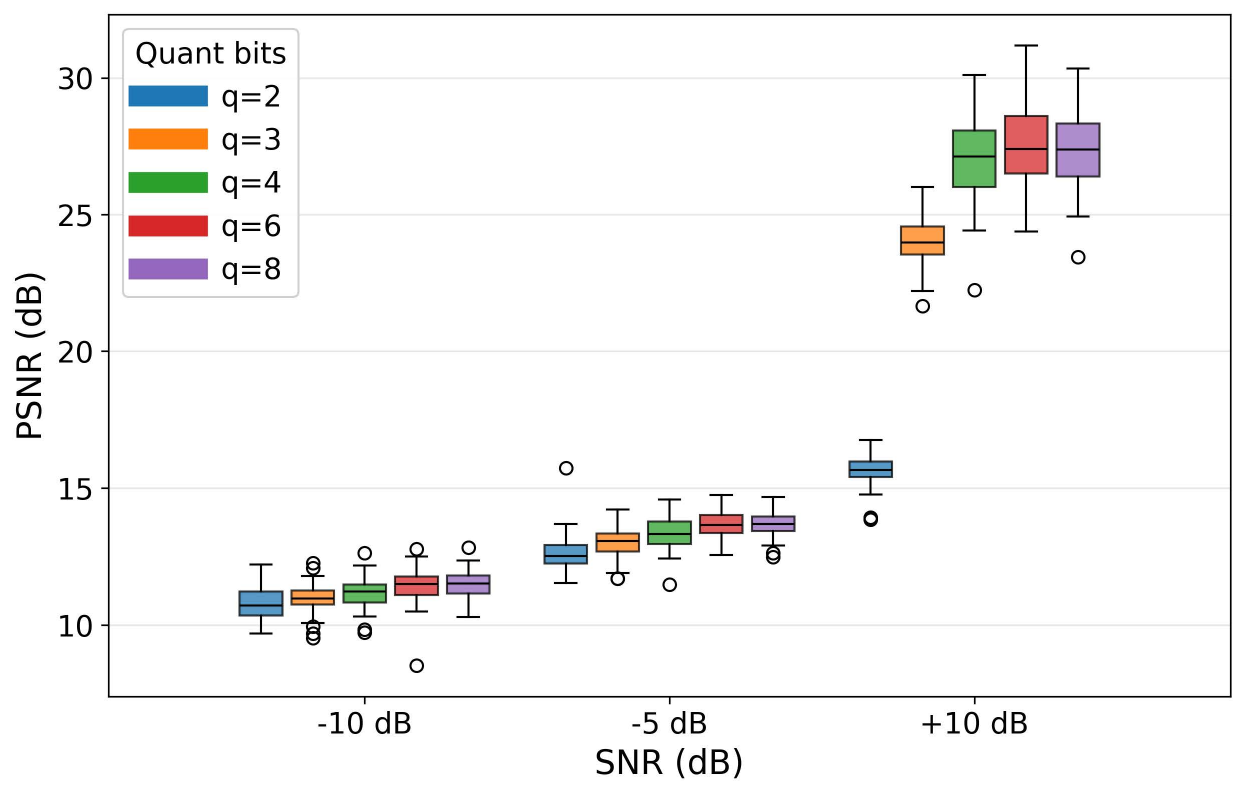}
        \caption{Reconstruction Quality (PSNR) (↑ better)}
    \end{subfigure}
    \begin{subfigure}[t]{0.33\linewidth}
        \centering
        \includegraphics[width=\linewidth]{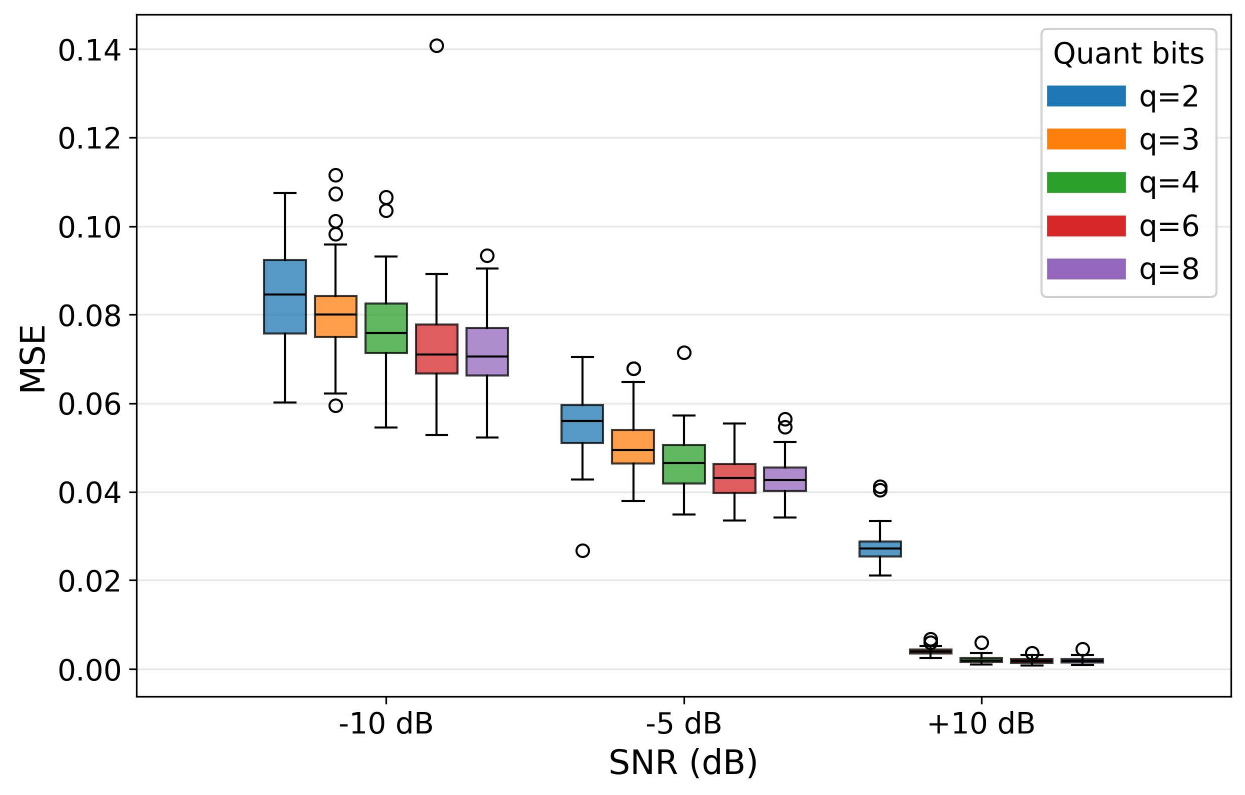}
        \caption{Pixel-Level Reconstruction Error (MSE)(↓ better)}
    \end{subfigure}
    \begin{subfigure}[t]{0.33\linewidth}
        \centering
        \includegraphics[width=\linewidth]{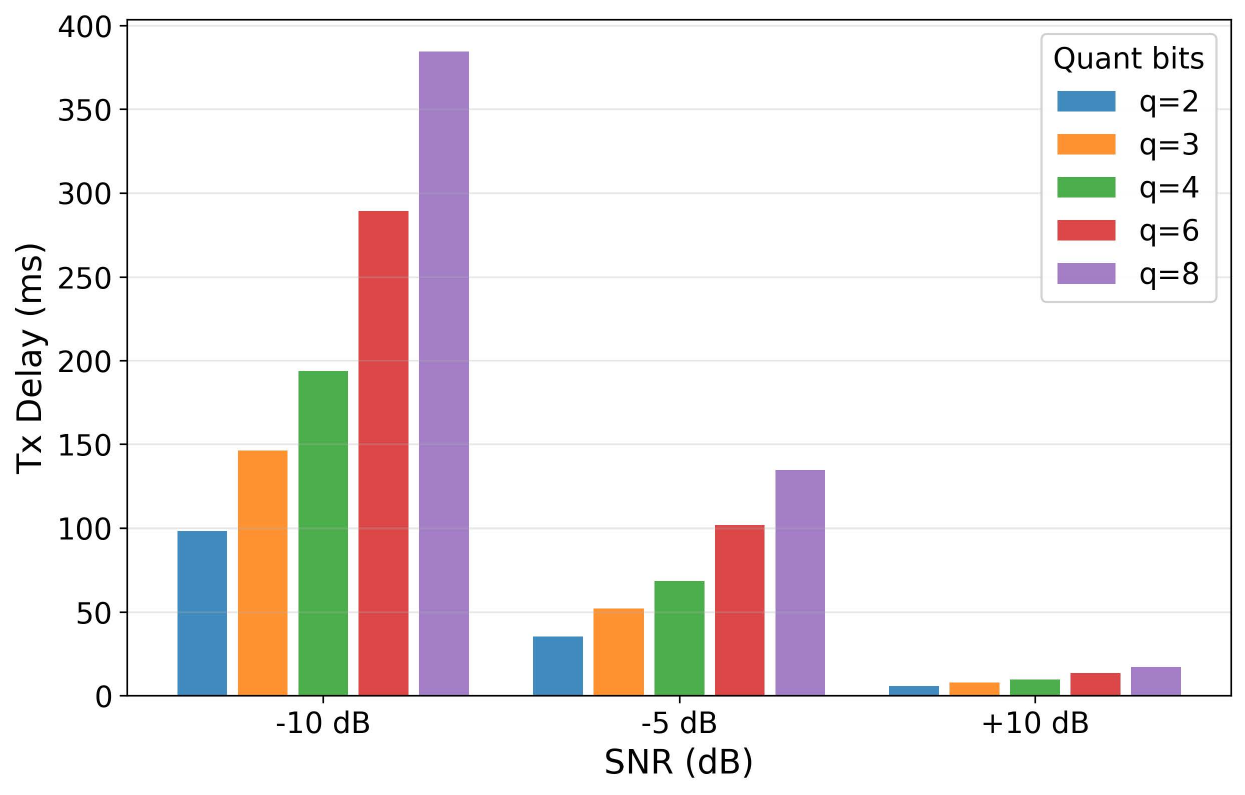}
        \caption{End-to-End Frame and Text Transmission Delay}
    \end{subfigure}
    \begin{subfigure}[t]{0.30\linewidth}
        \centering
        \includegraphics[width=\linewidth]{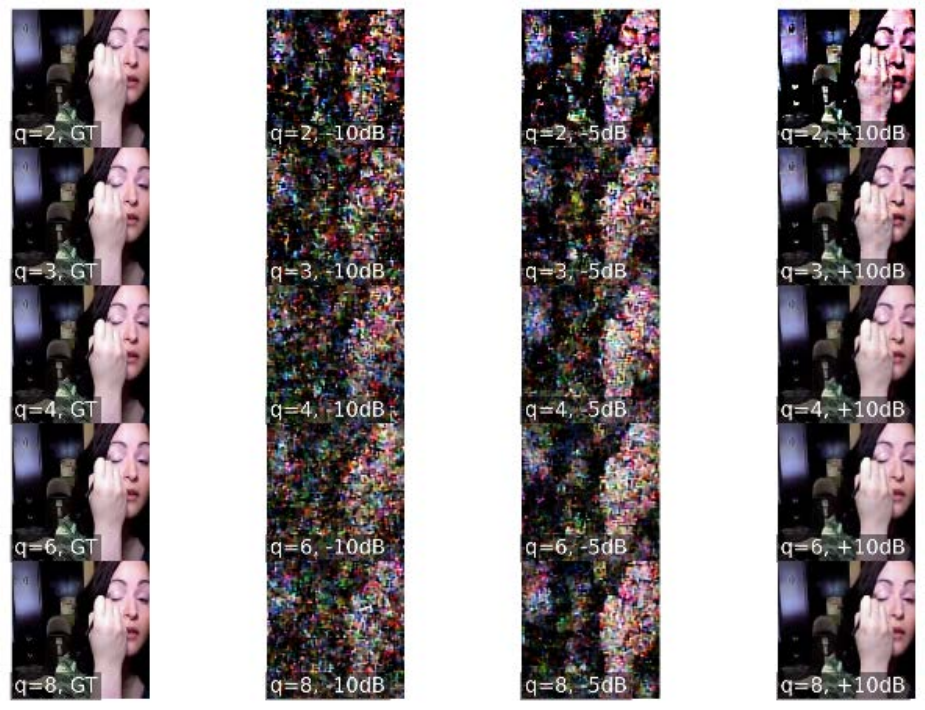}
        \caption{Visual Examples of Recovered Frames}
    \end{subfigure}

    \caption{Mirage with AE-based Semantic Communication under Varying Wireless Channel Conditions.}
    \label{fig:franklin-4sub}
\end{figure*}


\begin{figure*}[ht!]
    \centering
    \begin{subfigure}[t]{0.33\linewidth}
        \centering
        \includegraphics[width=\linewidth]{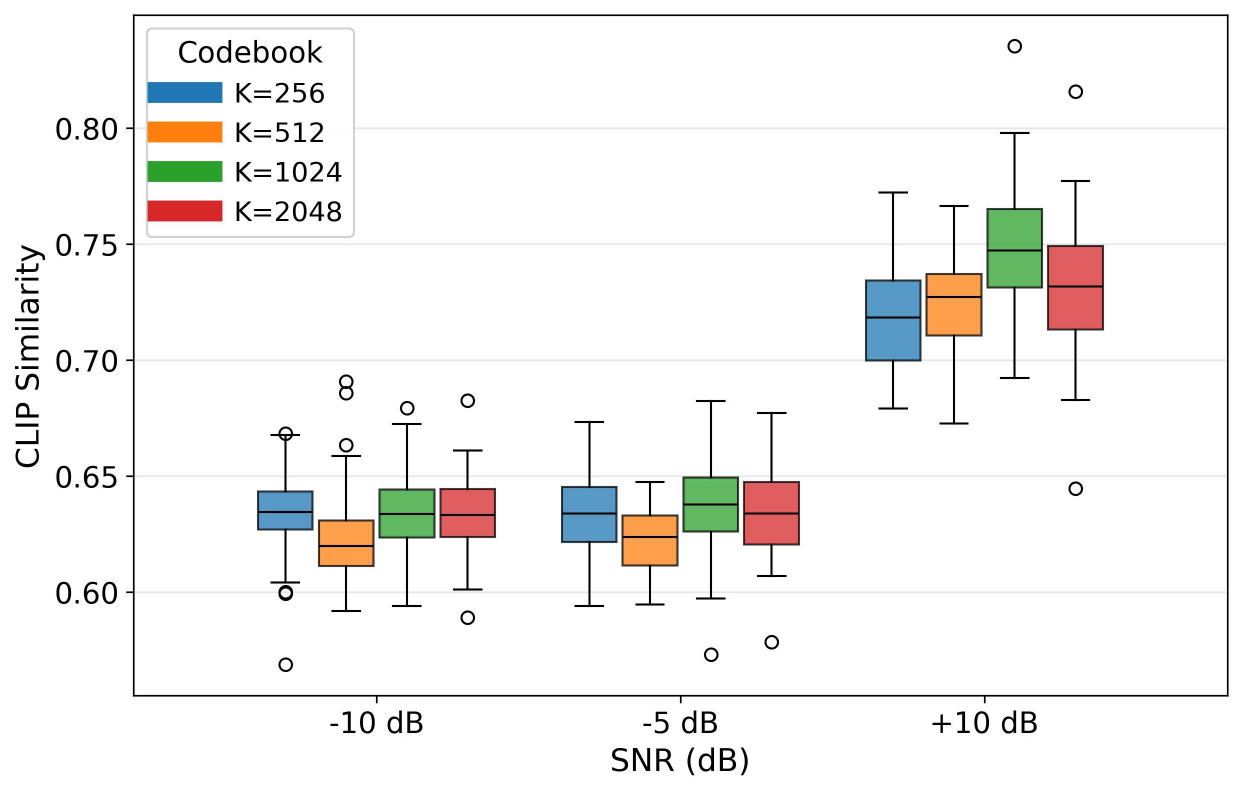}
        \caption{Semantic Similarity (CLIP) (↑ better)}
    \end{subfigure}
    \begin{subfigure}[t]{0.33\linewidth}
        \centering
        \includegraphics[width=\linewidth]{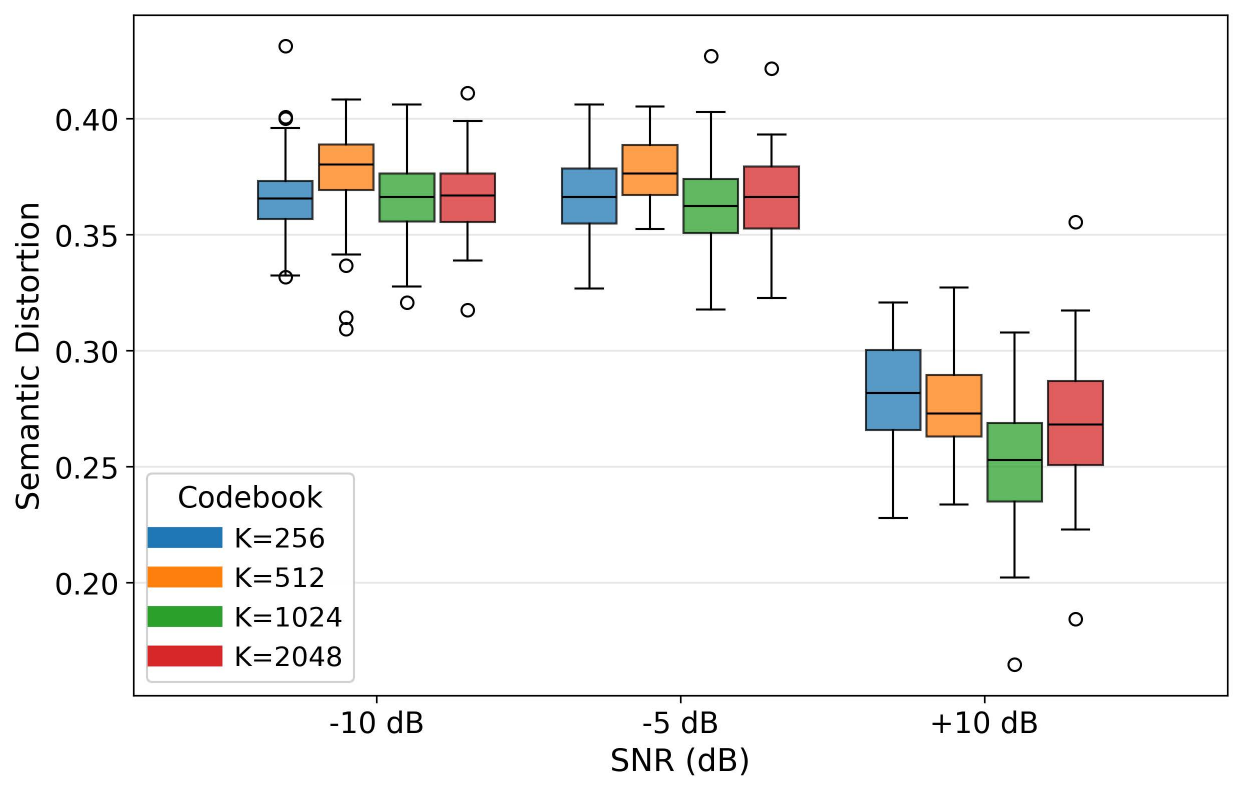}
        \caption{Semantic Distortion (↓ better)}
    \end{subfigure}
    \begin{subfigure}[t]{0.33\linewidth}
        \centering
        \includegraphics[width=\linewidth]{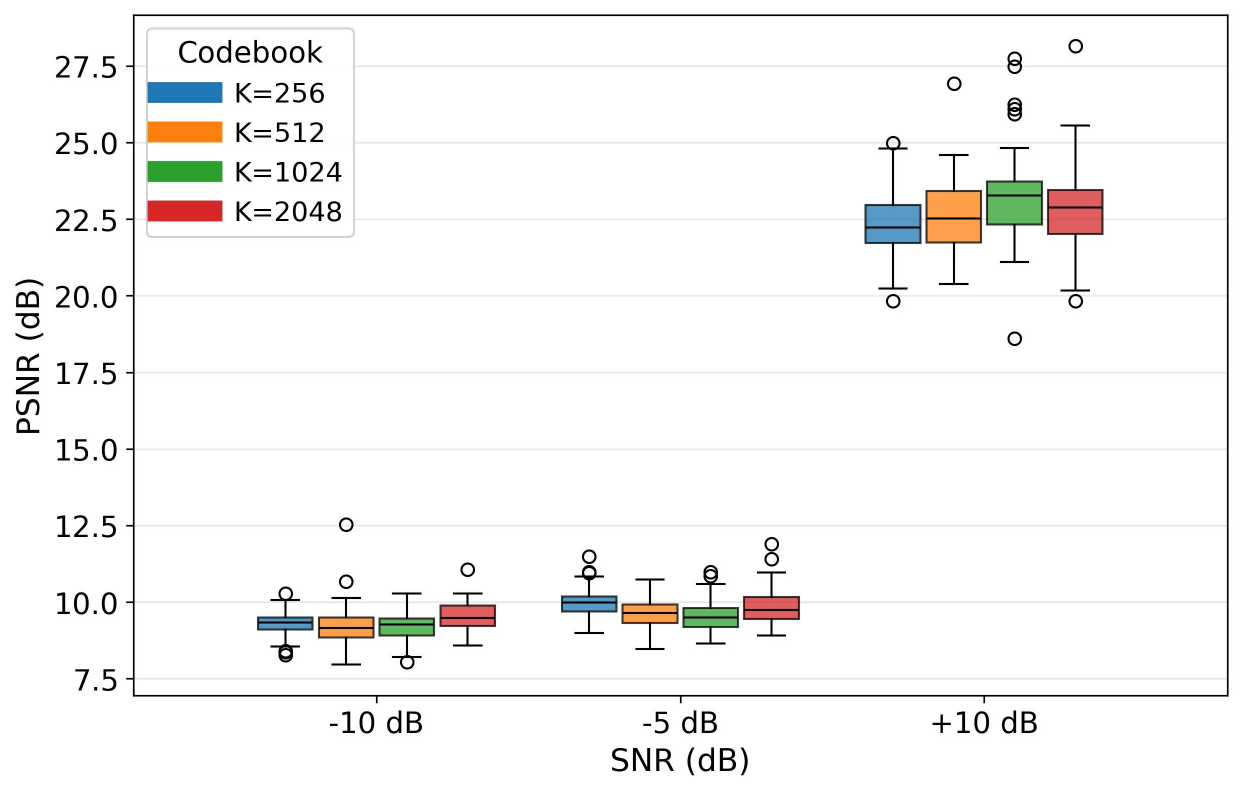}
        \caption{Reconstruction Quality (PSNR) (↑ better)}
    \end{subfigure}
    \begin{subfigure}[t]{0.33\linewidth}
        \centering
        \includegraphics[width=\linewidth]{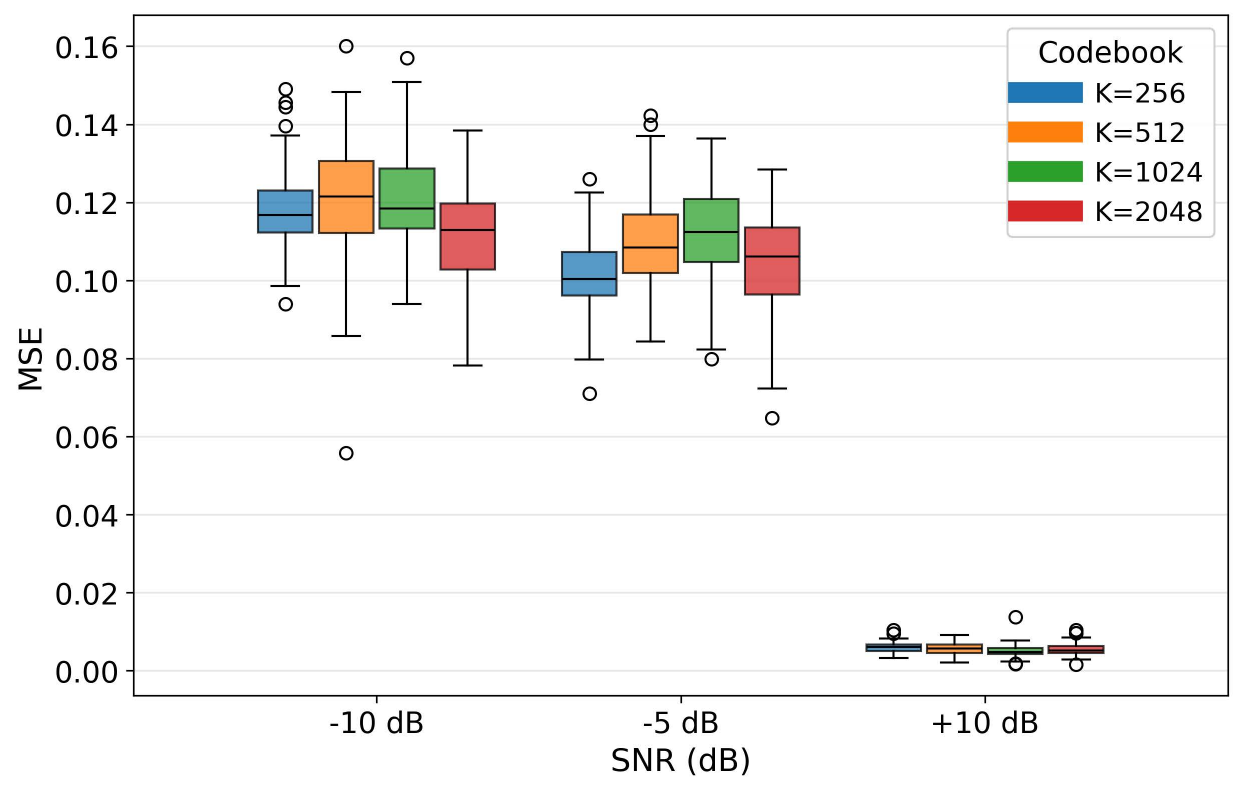}
        \caption{Pixel-Level Reconstruction Error (MSE)(↓ better)}
    \end{subfigure}
    \begin{subfigure}[t]{0.33\linewidth}
        \centering
        \includegraphics[width=\linewidth]{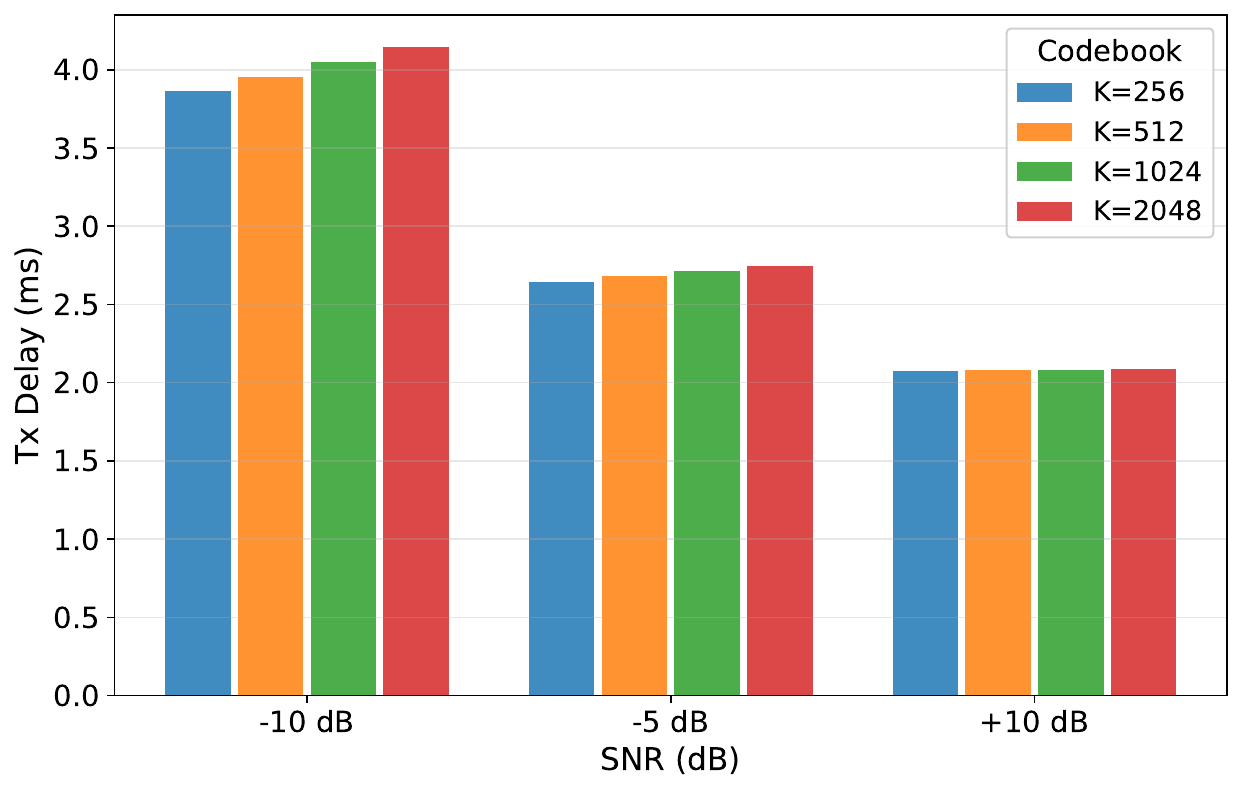}
        \caption{End-to-End Frame and Text Transmission Delay}
    \end{subfigure}
    \begin{subfigure}[t]{0.30\linewidth}
        \centering
        \includegraphics[width=\linewidth]{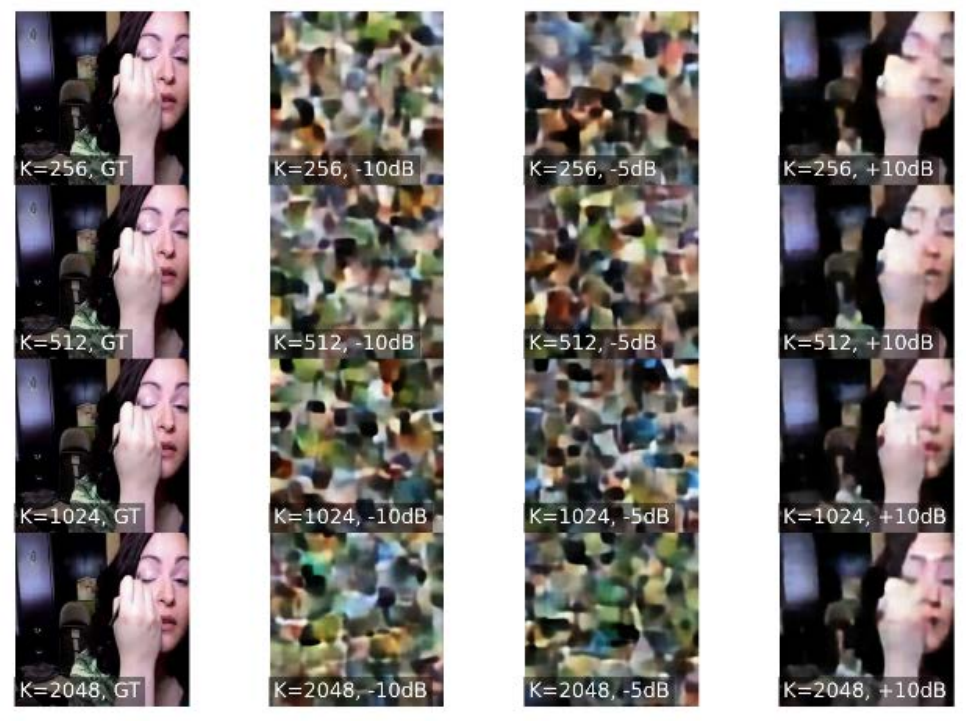}
        \caption{Visual Examples of Recovered Frames}
    \end{subfigure}

    \caption{Mirage with VQ-VAE-based Semantic Communication under Varying Wireless Channel Conditions.}
    \label{fig:franklin-4sub}
\end{figure*}

\subsection{Impact of Wireless Channel Conditions on Video Transmission Paradigms}

As SNR decreases, the resulting increase in BER leads to a more severe corruption of transmitted bits, which propagates into larger pixel-level reconstruction errors and ultimately affects perceptual and semantic quality. This mapping allows us to consistently study how wireless channel degradation impacts video transmission from bit-level errors to semantic degradation.

Figures~5--7 present a comparative evaluation of raw video transmission and Mirage-based semantic communication under varying channel conditions. As shown in Figure~5, raw video transmission is highly sensitive to channel quality. When SNR decreases from $10$~dB to $-10$~dB, the corresponding BER increases by several orders of magnitude, resulting in substantial pixel corruption. This effect is reflected in the sharp increase in pixel-level reconstruction error: the MSE rises from near zero at $10$~dB to above $0.15$ at $-10$~dB, while PSNR drops from over $30$~dB to below $10$~dB. Despite improvements in pixel fidelity at higher SNR, semantic similarity measured by CLIP remains low under poor channel conditions, indicating that accurate pixel reconstruction does not necessarily preserve semantic content.

Introducing semantic abstraction through Mirage significantly mitigates the impact of channel-induced bit errors. As shown in Figure~6, the AE-based semantic transmission achieves more stable semantic similarity and lower semantic distortion compared to raw transmission across different SNR levels. At $-10$~dB, AE-based transmission reduces MSE by more than $40\%$ relative to raw video transmission, while improving CLIP similarity by approximately $0.1$--$0.15$. However, because AE relies on continuous latent representations, its performance remains sensitive to quantization precision and channel noise, and semantic distortion still increases as SNR degrades.

Figure~7 further demonstrates the advantages of discrete semantic abstraction using VQ-VAE. Across all evaluated SNR conditions, Mirage with VQ-VAE-based transmission maintains consistently high semantic similarity and low semantic distortion, with CLIP similarity remaining within a narrow range of approximately $0.65$--$0.75$. Unlike raw and AE-based transmission, semantic quality becomes largely insensitive to channel variations once the discrete semantic indices are correctly received. Although pixel-level metrics such as PSNR and MSE no longer dominate, the generated frames preserve semantic structure and visual coherence, as confirmed by the qualitative examples. In addition, transmission latency is reduced by orders of magnitude compared to raw video transmission, highlighting the communication efficiency enabled by discrete semantic representations.


\begin{figure}[!t]
  \centering
\includegraphics[width=\linewidth]{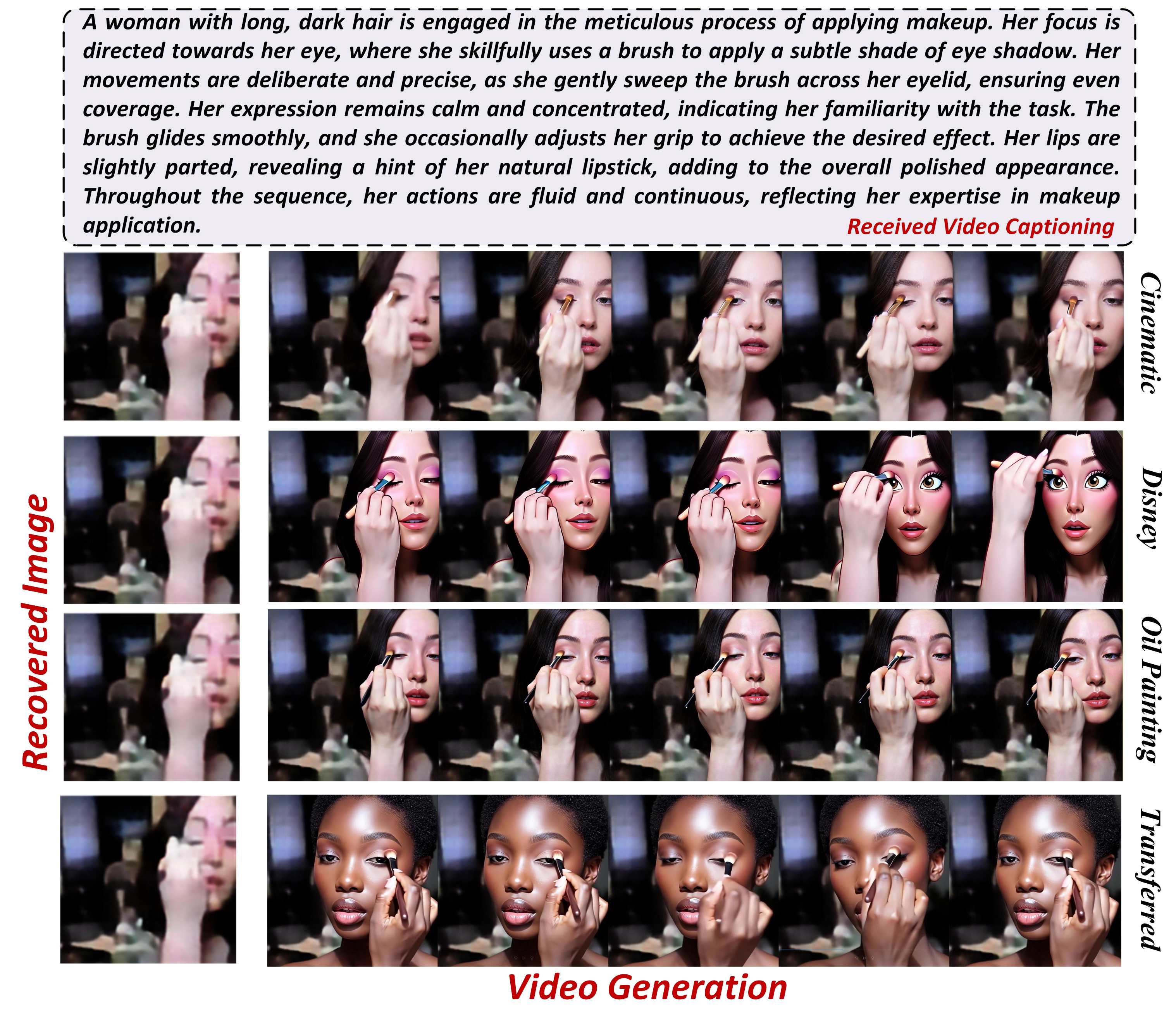}
  \caption{Receiver-Side Personalized Video Generation from Semantically Reconstructed Video ($\approx$22 dB PSNR) under Diverse Style and Identity Controls.}
  \Description{XXX.}
  \label{fig:VQVAE_recover}
\end{figure}

\subsection{Personalization and Controllability}

\begin{figure}[!t]
  \centering
  \includegraphics[width=\linewidth]{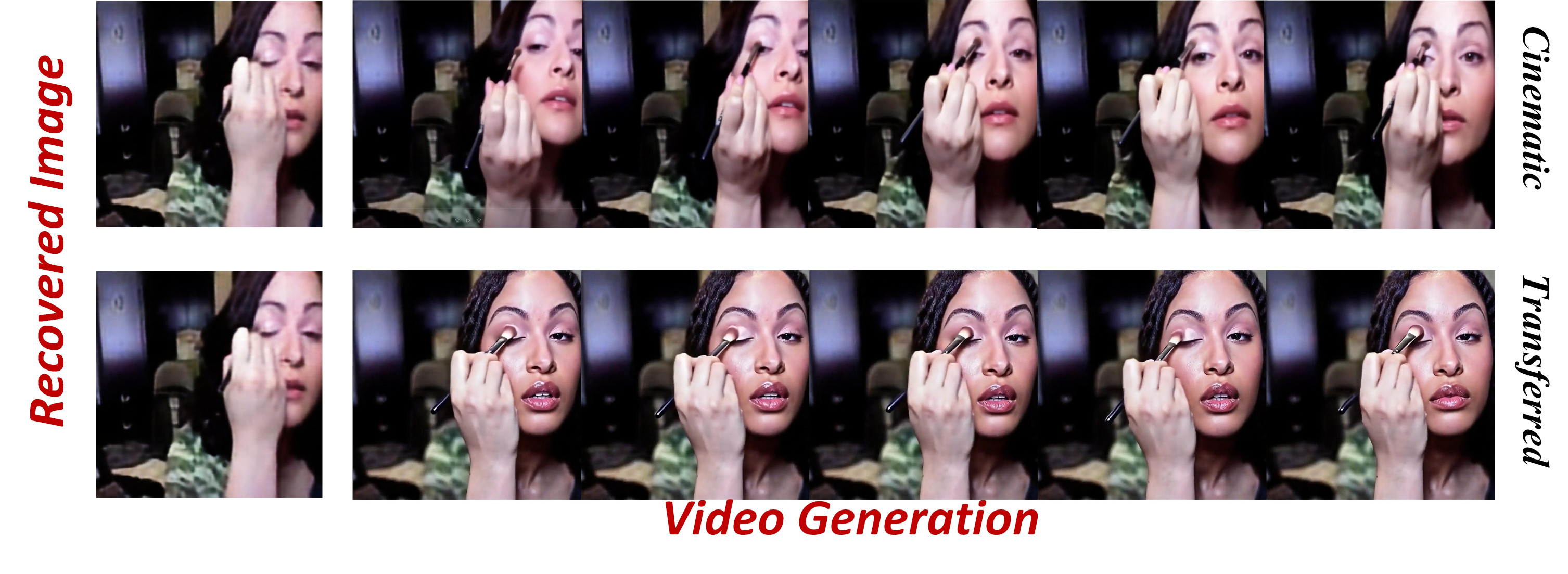}
  \caption{Receiver-side personalized video generation from Semantically Reconstructed video ($\approx$28 dB PSNR) under different style and identity settings.}
  \Description{XXX.}
  \label{fig:AE_recover}
\end{figure}

Figures \ref{fig:VQVAE_recover} and \ref{fig:AE_recover} jointly present personalized video generation results driven by semantically reconstructed video frames obtained via two representative semantic transmission schemes: VQ-VAE-based aggressive compression ($\approx$22 dB PSNR) and AE-based moderate compression ($\approx$28 dB PSNR). In both cases, original video frames are first transmitted in the semantic domain and reconstructed at the receiver—optionally with spatial resolution enhancement—before being used, together with received video captions, as conditioning inputs for downstream video generation. Despite the significantly lower pixel-level reconstruction quality of the VQ-VAE scheme, the reconstructed frames retain sufficient structural and semantic information, such as facial layout, pose, and motion continuity. As a result, style-oriented personalization (e.g., cinematic, Disney-style, and oil painting) remains visually coherent, with the generative model effectively enhancing contours, global appearance, and stylistic attributes even when fine-grained textures are partially lost. This demonstrates that high-fidelity pixel reconstruction is not a strict requirement for effective video generation when semantic consistency is preserved.

In contrast, the AE-based reconstruction provides higher visual fidelity and stronger consistency with the original frames, which directly benefits fidelity-oriented personalization. In the cinematic setting generated videos closely align with reconstructed content in appearance and temporal dynamics, illustrating how improved reconstruction quality can further enhance realism when bandwidth permits. Both transmission schemes support identity-transferred generation, where facial identity is intentionally replaced while preserving motion dynamics and semantic structure. This capability highlights a key advantage of semantic communication: identity attributes can be selectively altered or removed at the receiver to meet privacy or application-specific requirements, without affecting task-relevant semantics.

Consequently, these results indicate a clear trade-off between reconstruction fidelity and transmission efficiency. Crucially, the results confirm that compact semantic representations combined with receiver-side generative models and textual guidance are sufficient to support various personalization objectives, including style transformation, resolution enhancement, and privacy-preserving identity substitution. This validates Mirage’s core design principle: video communication should prioritize semantic and perceptual utility over exact signal fidelity. Even under aggressive semantic compression, Mirage preserves more than 90\% semantic similarity at the caption level, highlighting the sufficiency of semantic representations for downstream video generation.

\subsection{Semantic Preservation from Raw Video to Generated Video}

\begin{table}[t]
\centering
\caption{Semantic consistency between the generated video and the raw video was evaluated using BERTScore and SBERT cosine similarity.}

\label{tab:semantic_metrics}
\begin{adjustbox}{width=0.45\textwidth}
\begin{tabular}{lcccc}
\toprule
\textbf{Method} & \textbf{BERT-P} & \textbf{BERT-R} & \textbf{BERT-F1} & \textbf{SBERT} \\
\midrule
Cinematic-(AE)        & 0.786 & 0.788 & 0.787 & 0.959 \\
Cinematic-(VQVAE)     & 0.734 & 0.731 & 0.733 & 0.934 \\
Disney-(VQVAE)        & 0.718 & 0.727 & 0.722 & 0.875 \\
Oil Painting-(VQVAE)  & 0.753 & 0.764 & 0.758 & 0.917 \\
Transferred-(AE)      & 0.770 & 0.776 & 0.773 & 0.949 \\
Transferred-(VQVAE)   & 0.767 & 0.758 & 0.762 & 0.899 \\
\bottomrule
\end{tabular}
\end{adjustbox}
\end{table}

\begin{figure}[!t]
  \centering
  \includegraphics[width=\linewidth]{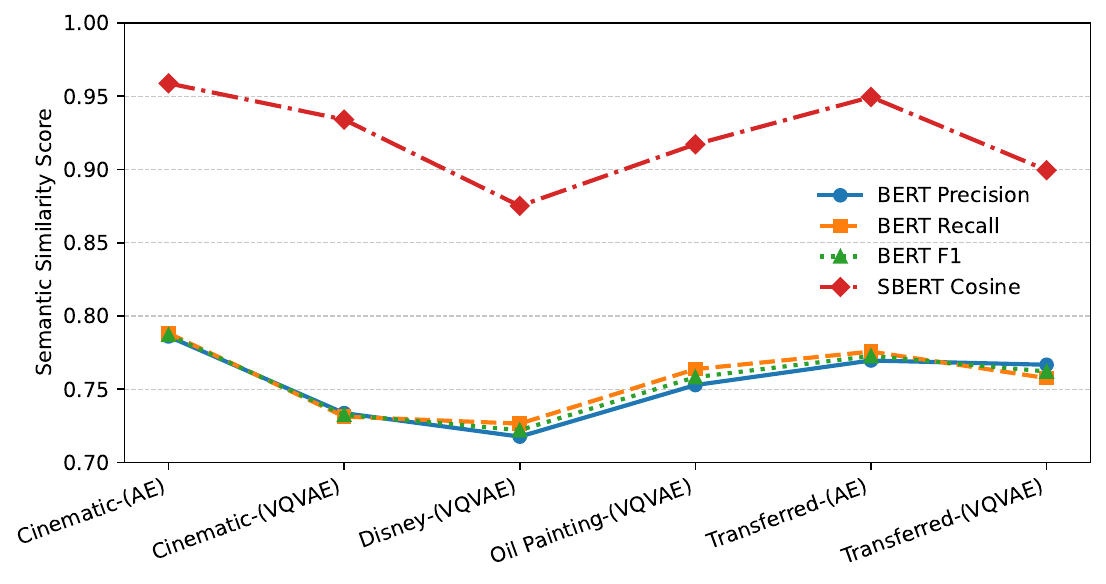}
  \caption{Evaluate the semantic consistency between the generated video and the raw video.}
  \Description{XXX.}
  \label{fig:video_semantic}
\end{figure}
 
Semantic consistency between the original and generated videos is evaluated by comparing their video captions using BERTScore and SBERT cosine similarity, as reported in Table~\ref{tab:semantic_metrics} and Figure \ref{fig:video_semantic}. In general, all configurations achieve high semantic alignment, with BERT-F1 scores ranging from 0.72 to 0.79 and SBERT cosine similarity consistently above 0.87, indicating strong preservation of high-level video semantics after semantic transmission and receiver-side generation.

AE-based reconstruction exhibits the highest semantic consistency, with Cinematic-(AE) achieving a BERT-F1 of 0.787 and an SBERT similarity of 0.959, benefiting from its higher reconstruction quality ($\approx$28 dB PSNR). In comparison, VQ-VAE-based transmission—despite operating at a lower reconstruction fidelity ($\approx$22 dB PSNR)—still maintains competitive semantic alignment. For instance, Cinematic-(VQVAE) reaches a BERT-F1 of 0.733 and SBERT similarity of 0.934, while Oil Painting-(VQVAE) achieves a BERT-F1 of 0.758 and SBERT similarity of 0.917, demonstrating robustness to aggressive semantic compression.

Notably, even under identity-transferred generation, where facial identity is intentionally altered, semantic similarity remains high (Transferred-(VQVAE): BERT-F1 0.762, SBERT 0.899). This confirms that while appearance attributes may change, motion dynamics and high-level semantic content are preserved. These results further validate that semantic communication prioritizes meaning preservation over pixel-level fidelity, enabling diverse personalization objectives without compromising semantic consistency.




\section{Conclusion}
In this paper, we introduce Mirage, a novel communication framework that redefines video transmission as a semantic and generative process, reducing communication costs while preserving perceptual quality. By leveraging compact semantic representations and generative reconstruction, Mirage offers a new approach for communication systems that focuses on semantic consistency rather than pixel-level fidelity. Mirage has significant implications for future communication networks, including 6G, next-generation Wi-Fi, and low-altitude systems where bandwidth, privacy, and personalization are critical. While not the final solution, we hope Mirage serves as a step toward more sustainable and flexible communication designs, demonstrating the potential of generative AI in networking.

\noindent\textbf{Ethics}: This work uses publicly available datasets that contain human faces. All datasets used in this paper are open-source and comply with their
respective licenses and usage policies. The generated faces shown in the evaluation are fully anonymized and synthetically reconstructed by generative models, where no identity information can be recovered from the original data. These visual examples are included solely to demonstrate generation quality and semantic consistency, rather than to represent real individuals.

\bibliographystyle{ACM-Reference-Format}
\bibliography{sample-base}

\end{document}